\documentclass[iop]{emulateapj}
\usepackage{fixltx2e}

\newcommand{\OII}{\mbox{O\hspace{0.25em}{\sc ii}}}

\usepackage{fixltx2e}
\usepackage{url}
\usepackage{amsmath}
\usepackage{hyperref}

\shorttitle{SLSNe in LSST}
\shortauthors{Villar, Nicholl \& Berger}

\begin{document}
\title{Superluminous Supernovae in LSST:  Rates, Detection Metrics, and Light Curve Modeling}

\author{V.~Ashley~Villar\altaffilmark{1}, Matt~Nicholl\altaffilmark{1}, and Edo~Berger\altaffilmark{1}}
\affil{Harvard-Smithsonian Center for Astrophysics, 60 Garden Street, Cambridge, Massachusetts 02138, USA; \href{mailto:vvillar@cfa.harvard.edu}{vvillar@cfa.harvard.edu}}

\begin{abstract}
We explore and demonstrate the capabilities of the upcoming Large Synoptic Survey Telescope (LSST) to study Type I superluminous supernovae (SLSNe). We first fit the light curves of 58 known SLSNe at $z\approx 0.1-1.6$, using an analytical magnetar spin-down model implemented in {\tt MOSFiT}. We then use the posterior distributions of the magnetar and ejecta parameters to generate thousands of synthetic SLSN light curves, and we inject those into the LSST Operations Simulator (OpSim) to generate realistic $ugrizy$ light curves. We define simple, measurable metrics to quantify the detectability and utility of the light curve, and to measure the efficiency of LSST in returning SLSN light curves satisfying these metrics.  We combine the metric efficiencies with the volumetric rate of SLSNe to estimate the overall discovery rate of LSST, and we find that $\approx 10^4$ SLSNe per year with $>10$ data points will be discovered in the Wide-Fast-Deep (WFD) survey at $z\lesssim3.0$, while only $\approx 15$ SLSNe per year will be discovered in each Deep Drilling Field at $z\lesssim4.0$. To evaluate the information content in the LSST data, we refit representative output light curves with the same model that was used to generate them.  We correlate our ability to recover magnetar and ejecta parameters with the simple light curve metrics to evaluate the most important metrics. We find that we can recover physical parameters to within 30\% of their true values from $\approx18$\% of WFD light curves. Light curves with measurements of both the rise and decline in $gri$-bands, and those with at least fifty observations in all bands combined, are most information rich, with $\approx 30$\% of these light curves having recoverable physical parameters to $\approx30$\% accuracy. WFD survey strategies which increase cadence in these bands and minimize seasonal gaps will maximize the number of scientifically useful SLSN light curves. Finally, although the Deep Drilling Fields will provide more densely sampled light curves, we expect only $\approx 50$ SLSNe with recoverable parameters in each field in the decade-long survey.
\end{abstract}


\section{Introduction}

Type I Superluminous supernovae (SLSNe) are an observationally-classified class of transients that typically reach a peak absolute magnitude of $\lesssim -20$ mag and display unique early-time spectra with \OII\ absorption superposed on a hydrogen-free blue continuum \citep{chomiuk2011pan,quimby2011hydrogen,gal2012luminous}. These events also typically exhibit long durations, with a time to rise and decline by one magnitude of $t_\mathrm{dur}\gtrsim 50$ days, allowing them to radiate $\approx 10^{51}$ erg in the optical/UV, comparable to the kinetic energies of normal core-collapse SNe. Despite their high luminosities and long durations, SLSNe are a relatively recent discovery due to the advent of untargeted wide-field time-domain surveys.  These surveys are essential due to the low volumetric rate and low-luminosity host galaxies of SLSNe \citep{neill2010extreme,lunnan2014,chen2017superluminous,leloudas2015spectroscopy,angus2016hubble,perley2016host,schulze2018cosmic}. 

There is ongoing debate about the energy source of SLSNe. Unlike hydrogen-rich Type II SLSNe which appear to be powered by interaction with a dense circumstellar medium \citep{chevalier2011shock}, such interaction is disfavored as the dominant heating source for Type I SLSNe, due to the exceptionally large CSM mass required to reproduce the bright observed light curves \citep{moriyareview2018}, coupled with low-density environments suggested by X-ray \citep{margutti2017results} and radio \citep{nicholl2016sn,coppejans2018jets} observations. Instead, a central engine model is preferred, and it appears to explain the light curve shapes and diversity \citep{nicholl2017magnetar}, early-time spectra (e.g., \citealt{dessart2012superluminous,howell2013two,mazzali2016}), and the velocity and density structures inferred from nebular spectra  \citep{nicholl2016superluminous,jerkstrand2017long,nicholl2018nebular}.

Currently, the best central engine candidate to power SLSNe is a rapidly spinning {\it magnetar}, or a pulsar with a strong magnetic field ($B\gtrsim10^{13}$ G; \citealt{kasen2010supernova,woosley2010bright,inserra2013super,chatzopoulos2013analytical,metzger2015diversity,nicholl2017magnetar}). The magnetar model can  explain the diversity of SLSN light curves \citep{nicholl2017magnetar,villar2017theoretical}, and the inferred velocities and temperatures \citep{moriyareview2018}. Recently, \citealt{nicholl2017magnetar} fit a sample of 38 well-observed SLSNe with a semi-analytical magnetar model and found that this model is able to reproduce the observed light curves using a fraction of parameter space (see also, \citealt{nicholl2015diversity,prajs2016volumetric,yu2017statistical,liu2017monte}). 

Statistical population studies like these are essential for mapping the properties of SLSNe. Currently, about 10 SLSNe are discovered per year \citep{guillochon2017open}\footnote{See \url{sne.space}.}, and this low rate allows for detailed spectroscopic follow-up of each event. However, future surveys will lead to a substantial increase in the discovery rate.  For example, \citet{tanaka2012detectability} and \citet{tanaka2013detectability} explored several future optical and near-infrared (NIR) surveys, concluding that missions like WFIRST and LSST will find $\sim 10^2-10^4$ SLSNe per year. Due to limited spectroscopic resources, it is essential to explore what information can be obtained about these large samples from light curves alone; namely, their diverse observational properties \citep{nicholl2015diversity}, progenitor populations \citep{lunnan2014hydrogen}, host galaxies \citep{berger2012ultraluminous,schulze2018cosmic}, and potentially cosmological parameters \citep{inserrasmart2014,scovacricchi2015cosmology}.

Here, we explore and study the characteristics of SLSNe observed by the upcoming Large Synoptic Survey Telescope (LSST), an 8.4-m diameter telescope with a 9.6 deg$^2$ field-of-view that will conduct several 10-year wide-field surveys across the Southern hemisphere in the $ugrizy$ filters. The current LSST observing strategy spends the majority ($\gtrsim 90\%$) of the time executing the Wide-Fast-Deep (WFD) survey, covering 18,000 deg$^2$ with a cadence of roughly four days in any filter and ten days in a specific filter, and with a per-visit limiting magnitude of $m_\mathrm{lim,gr}\approx 24.5$ mag. About 5\% of the observing time will focus on several Deep Drilling fields (DDFs), each comprised of a single pointing with five times more cumulative imaging than a typical field in the WFD survey and with a nightly stack limiting magnitude of $m_\mathrm{lim,gr}\approx 26.5$ mag.  The remaining time will be split between the Galactic Plane and South Celestial Pole surveys. We focus on the WFD survey and DDFs in this study (see \citealt{lsstsciencecollab} for technical details). 

The large survey area, cadence and depth make LSST a potential powerhouse for time-domain astronomy, particularly in the case of volumetrically-rare events like SLSNe.  However, there are two key questions that must be addressed to maximize the potential of LSST. First, it is essential to quantify the number of SLSNe that LSST will discover, their redshift distribution, and their observational properties. Second, it is vital to predict the information content from LSST light curves alone to account for cases which will lack spectroscopic follow up. In this work we perform the first detailed study of SLSNe discovered with LSST using an observationally-motivated suite of SLSN models and a realistic observational simulator provided by the LSST collaboration.

The paper is structured as follows. In \S\ref{sec:sim}, we outline the simulations used to produce realistic SLSN light curves as they would appear in the LSST WFD survey and DDFs. In \S\ref{sec:char}, we describe the characteristics of the SLSNe discovered by LSST based on our simulation results. In \S\ref{sec:recover}, we discuss the ability to recover physical parameters from the LSST light curves, and quantify the information content of the simulated light curves. We conclude in \S\ref{sec:conclusions}.  All magnitudes are reported in the AB system, and we assume a standard cosmology, with $H_0 = 67.7$ km s$^{-1}$ Mpc$^{-1}$, $\Omega_M = 0.307$, and $\Omega_{\Lambda} = 0.691$ \citep{planck2016astronomy}.

\begin{figure}[t!]
\includegraphics[width=0.48\textwidth]{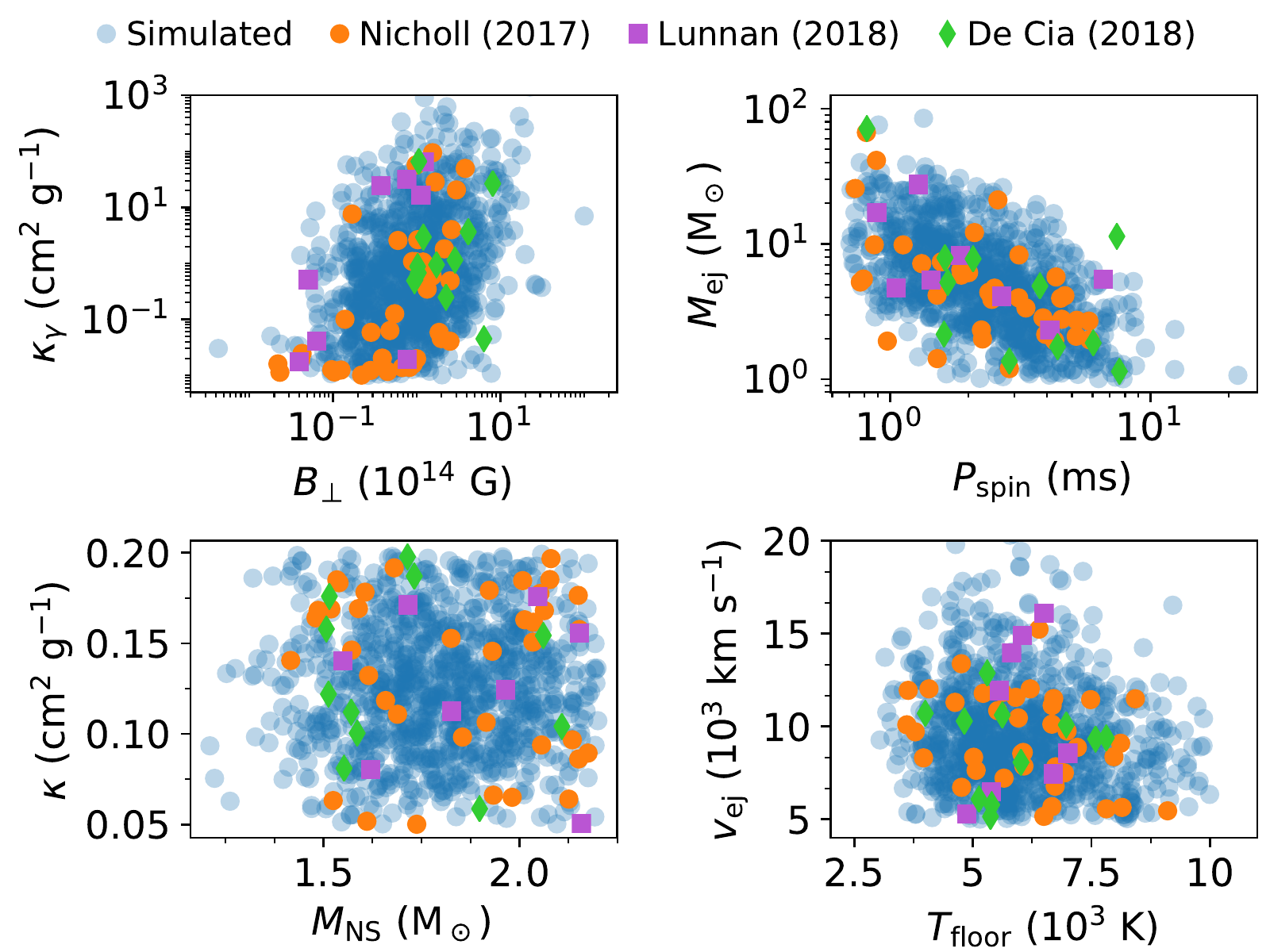}
\centering
\caption{A subset of the model parameters for the SLSN magnetar model, showing the results for the observed sample (orange circles: \citealt{nicholl2017magnetar}; purple squares: \citealt{lunnan2018}; green diamonds; \citealt{decia2017}), and our simulated population (blue circles). The simulated sample captures intrinsic correlations between parameters (including nuisance parameters) that are seen within the observed population.} 
\label{fig:params}
\end{figure}

\section{Simulation Set-Up}\label{sec:sim}

To simulate the observable SLSN population in the LSST surveys we construct a sample of light curve models and inject these into the LSST Operations Simulator ({\tt OpSim}). We describe these steps in the following subsections. 

\subsection{Constructing Simulated SLSN Light Curves}

We construct a sample of simulated light curves based on known events from the literature.  Several studies have previously aggregated SLSN light curves (e.g. \citealt{nicholl2015diversity,2017MNRAS.464.3568P,2017ApJ...845...85L,decia2017,lunnan2018}).  Recently, \citet{nicholl2017magnetar} uniformly modeled a sample of 38 SLSNe requiring the events to be spectroscopically classified and to have some photometric data near peak. As such, the sample spans a range of peak luminosities and light curve timescales. Here we combine the sample of \citet{nicholl2017magnetar} with 12 events discovered by the Palomar Transient Factory (PTF; \citealt{decia2017}) and 8 events discovered in the Pan-STARRS1 Medium Deep Survey (PS1-MDS; \citealt{lunnan2018}), leading to 58 spectroscopically classified SLSNe spanning a wide range of observational properties (see Table~\ref{table:events}). 

We model the 21 PTF and PS1-MDS SLSNe with the same model described in \citet{nicholl2017magnetar}. In short, we use the open-source code {\tt MOSFiT} \citep{guillochon2018mosfit} to fit a magnetar spin-down model to the multi-band light curves. We assume a modified blackbody spectral energy distribution (SED) in which flux is linearly suppressed below a ``cutoff" frequency of 3000\AA. This SED shape is consistent with observed SEDs (\citealt{chomiuk2011pan,nicholl2016extreme}, but see also \citealt{yan2018} who argue that more UV variation is seen in SLSN SEDs). We remove data in which a pre-peak ``bump" is observed (see \citealt{nicholl2016seeing}).  Our best-fit parameters and their $1\sigma$ error bars are provided in Table~\ref{table:results1}. 

We use the sample of 58 fitted events to generate our simulated SLSN light curves. Because drawing walkers directly from the model posteriors would lead to undersampling of the parameter space, we sample from a model of the underlying population distribution as follows. We draw one walker from the posterior of each event to create a distribution of the model parameters of the underlying population. We fit this distribution to a truncated multivariate log-Gaussian which allows us to capture the correlations between parameters observed in the sample events. We place physically-motivated limits on the parameters, as listed in Table~\ref{table:lims}. We then draw samples from the population distribution to generate 1,000 events per redshift bin of $\Delta z=0.1$ from $z=0.1$ to $z=6.0$. Finally, from this sample we eliminate events with $M_\mathrm{r}>-20$ mag; although these magnetar-powered events may exist in nature, they are not necessarily distinguishable from the broader population of Type I SNe. The parameters for the modeled SLSNe and for our simulated events are shown in Figure~\ref{fig:params}.

\begin{deluxetable*}{lllllll}
\tablecolumns{7} 
\tablewidth{0pt} 
\tablecaption{Type I SLSNe Used in this Analysis \label{table:events}} 
\tablehead{ 
\colhead{Name}& \colhead{Redshift} & \colhead{Reference} & &\colhead{Name}& \colhead{Redshift} & \colhead{Reference}}
\startdata 
PTF11hrq	&	0.057	&	\citet{decia2017}				&	&	PTF09cwl*	&	0.350	&	\citet{quimby2011hydrogen}\\
PTF10hgi*	&	0.099	&	\citet{inserra2013super}		&	&				& 			&	\citet{decia2017}\\
			&			&	\citet{decia2017}				&	&	SN2006oz*	&	0.376	&	\citet{leloudas2012sn}\\
Gaia16apd*	&	0.102	&	\citet{yan2017far}				& 	&	PTF13cjq	&	0.396	&	\citet{decia2017}\\
			&			&	\citet{nicholl2017ultraviolet}	& 	&	PTF13bdl	&	0.403	&	\citet{decia2017}\\
			&			&	\citet{kangas2017gaia16apd}		& 	&	iPTF13dcc*	&	0.431	&	\citet{vreeswijk2017early}\\
PTF12hni	&	0.106	&	\citet{decia2017}				& 	&				& 			&	\citet{decia2017}\\
PTF12dam*	&	0.107	&	\citet{nicholl2013slowly}		&	&	PTF10vqv	&	0.452	&	\citet{decia2017}\\
			&			&	\citet{chen2015host}			& 	&	PTF09atu*	&	0.502	&	\citet{quimby2011hydrogen}\\
			&			&	\citet{decia2017}				& 	&				& 			&	\citet{decia2017}\\
			&			&	\citet{vreeswijk2017early}		&	&	PS1-14bj*	&	0.522	&	\citet{lunnan2016ps1}\\
SN2015bn*	&	0.114	&	\citet{nicholl2016sn}			& 	&				& 			&	\citet{lunnan2018}\\
			&			&	\citet{nicholl2016superluminous}&	&	PS1-12bqf	&	0.522	&	\citet{lunnan2018}\\
PTF10nmn	&	0.124	&	\citet{decia2017}				&	&	PS1-11ap*	&	0.524	&	\citet{mccrum2013superluminous}\\
SN2007bi*	&	0.128	&	\citet{gal2009supernova}		& 	&				& 			&	\citet{lunnan2018}\\
SN2011ke*	&	0.143	&	\citet{inserra2013super}		&	&	DES14X3taz*	&	0.608	&	\citet{smith2016des14x3taz}\\
			&			&	\citet{decia2017}				& 	&	PS1-10bzj*	&	0.650	&	\citet{lunnan2013ps1}\\
SSS120810*	&	0.156	&	\citet{nicholl2014superluminous}& 	&				& 			&	\citet{lunnan2018}\\
PTF10bfz	&	0.170	&	\citet{decia2017}				&	&	DES13S2cmm*	&	0.663	&	\citet{papadopoulos2015des13s2cmm}\\
SN2012il*	&	0.175	&	\citet{inserra2013super}		& 	&	PS1-11bdn	&	0.738	&	\citet{lunnan2018}\\
PTF12gty	&	0.177	&	\citet{decia2017}				& 	&	iPTF13ajg*	&	0.740	&	\citet{vreeswijk2014hydrogen}\\
PTF11rks*	&	0.192	&	\citet{inserra2013super}		&	&				&			&	\citet{decia2017}\\
			&			&	\citet{decia2017}				&	&	PS1-13gt	&	0.884	&	\citet{lunnan2018}\\
PTF10aagc	&	0.207	&	\citet{decia2017}				& 	&	PS1-10awh*	&	0.908	&	\citet{chomiuk2011pan}\\
iPTF15esb*	&	0.224	&	\citet{yan2017hydrogen}			& 	&				&			&	\citet{lunnan2018}\\
SN2010gx*	&	0.230	&	\citet{pastorello2010ultra}		&	&	PS1-10ky*	&	0.956	&	\citet{chomiuk2011pan}\\
			&			&	\citet{quimby2011hydrogen}		&	&				& 			&	\citet{lunnan2018}\\
			&			&	\citet{decia2017}				&	&	PS1-11aib	&	0.997	&	\citet{lunnan2018}\\
SN2011kf*	&	0.245	&	\citet{inserra2013super}		& 	&	PS1-10ahf*	&	1.1		&	\citet{mccrum2015selecting}\\
iPTF16bad*	&	0.247	&	\citet{yan2017hydrogen}			& 	&				& 			&	\citet{lunnan2018}\\
LSQ14mo*	&	0.253	&	\citet{chen2017evolution}		&	&	SCP-06F6*	&	1.190	&	\citet{barbary2008discovery}\\ 
LSQ12dlf*	&	0.255	&	\citet{nicholl2014superluminous}& 	&	PS1-10pm*	&	1.206	&	\citet{mccrum2015selecting}\\
			&			&	\citet{decia2017}				& 	&				& 			&	\citet{lunnan2018}\\
PTF09cnd*	&	0.258	&	\citet{quimby2011hydrogen}		&	&	PS1-11tt	&	1.283	&	\citet{lunnan2018}\\
SN2013dg*	&	0.265	&	\citet{nicholl2014superluminous}& 	&	PS1-11afv	&	1.407	&	\citet{lunnan2018}\\
PTF13bjz	&	0.271	&	\citet{decia2017}				& 	&	SNLS-07D2bv*&	1.50	&	\citet{howell2013two}\\  
SN2005ap*	&	0.283	&	\citet{quimby2007sn}			& 	&	PS1-13or	&	1.52	&	\citet{lunnan2018}\\
PTF10uhf	&	0.288	&	\citet{decia2017}				& 	&	PS1-11bam*	&	1.565	&	\citet{berger2012ultraluminous}\\  
PTF12mxx	&	0.327	&	\citet{decia2017}				& 	&	PS1-12bmy	&	1.572	&	\citet{lunnan2018}\\	
iPTF13ehe*	&	0.343	&	\citet{yan2015detection}		& 	&				& 			&	\citet{lunnan2018}\\ 
			&			&	\citet{decia2017}				&	&	SNLS-06D4eu*&	1.588	&	\citet{howell2013two}\\
LSQ14bdq*	&	0.345	&	\citet{nicholl2015lsq14bdq}		&	&				&			&	
\enddata 
\tablecomments{$^*$Best fit magnetar parameters presented in  \citealt{nicholl2017magnetar}. Other magnetar parameters presented in Table 3.} 
\end{deluxetable*}

\begin{figure}[t]
\includegraphics[width=0.48\textwidth]{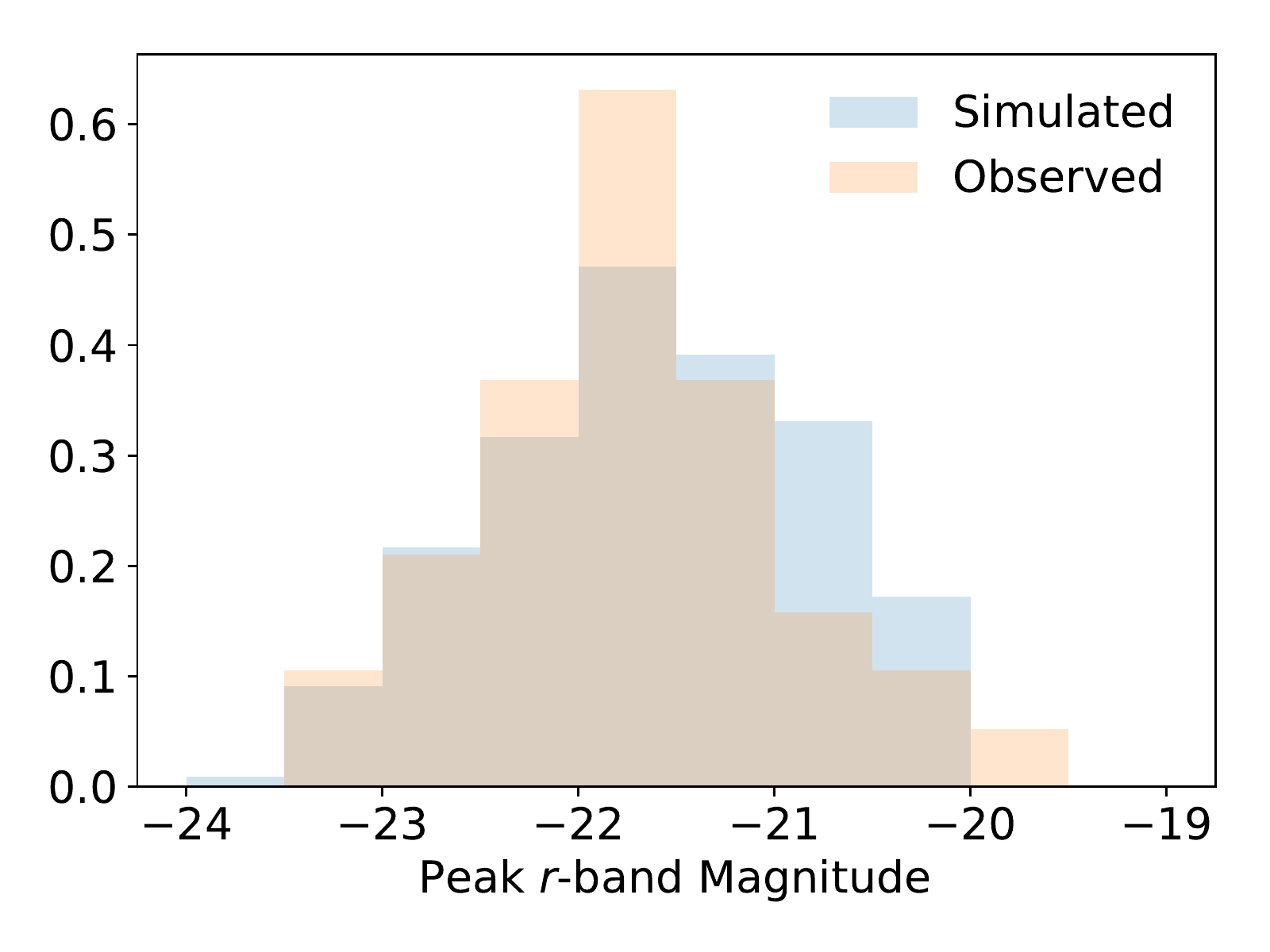}
\centering
\caption{Observed (orange) and simulated (blue) $r$-band peak luminosity function for SLSNe. The luminosity functions are in good agreement, with only $\approx 5$\% of our models extending to brighter $r$-band peak luminosities ($M_r\lesssim -23$) than currently observed.}
\label{fig:lumfunc}
\end{figure}

\begin{figure}[t]
\includegraphics[width=0.48\textwidth]{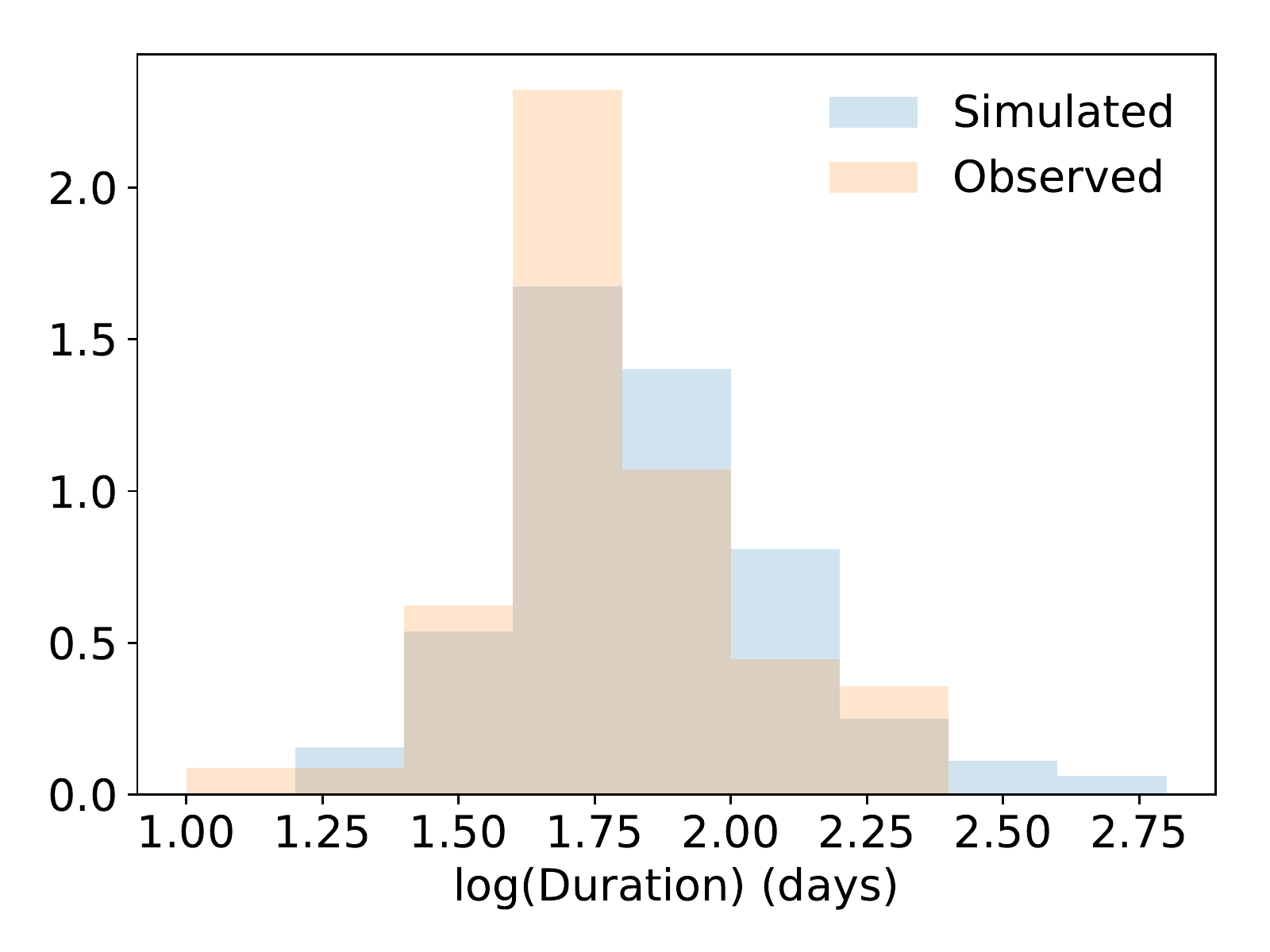}
\centering
\caption{Same as Figure~\ref{fig:lumfunc} but for the $r$-band duration ($t_\mathrm{dur}$) of the observed (orange) and simulated (blue) SLSNe. The duration distributions are in good agreement, with only $\approx 2$\% of our models extending to durations longer than those in our observed sample.}
\label{fig:durfunc}
\end{figure}

The resulting $r$-band peak-luminosity function of our simulated events and the observed sample are shown in Figure~\ref{fig:lumfunc}. Our simulated luminosity function is consistent with that derived by \citealt{nicholl2017magnetar}.  We do not attempt to correct for any potential observational biases within the various surveys, as we expect these effects to be small relative to the overall uncertainty in the volumetric rate (approximately a factor $\approx 3-5$; see \citealt{quimby2013rates,mccrum2015selecting,prajs2016volumetric} and \S\ref{sec:recover}). Similarly, we show the duration distribution of the known SLSNe and of our simulated events in Figure~\ref{fig:durfunc}, finding a good agreement between the two.

\begin{deluxetable}{ccc} 
\tablecolumns{3} 
\tablewidth{0pt} 
\tablecaption{Model Parameters and Imposed Limits \label{table:lims}} 
\tablehead{ 
\colhead{Parameter}& \colhead{Min} & \colhead{Max}}
\startdata 
$P_\mathrm{spin}$/ms & 0.7 & 100$^*$\\
$B_\bot/10^{14}$ G & $10^{-2*}$ & $10$\\
$M_\mathrm{ej}$/M$_\odot$ & 1 & 100$^*$\\
$v_\mathrm{ej}/10^4$ km s$^{-1}$ & 0.5 & 100$^*$\\
$\kappa$/ g cm$^{-2}$ & 0.05 & 0.2\\
$\kappa_\gamma$/ g cm$^{-2}$ & $10^{-2}$ & $10^3$\\
$M_\mathrm{NS}$/M$_\odot$ & 1.4 & 2.2\\
$T_\mathrm{floor}/10^3$ K & 0.1$^*$ & 50$^*$
\enddata 
\tablecomments{Parameters described in detail in \citealt{nicholl2017magnetar}.\\$^*$Indicates limits well within the tail of the Gaussian distribution.} 
\end{deluxetable}

\begin{figure*}[t]
\includegraphics[width=\textwidth]{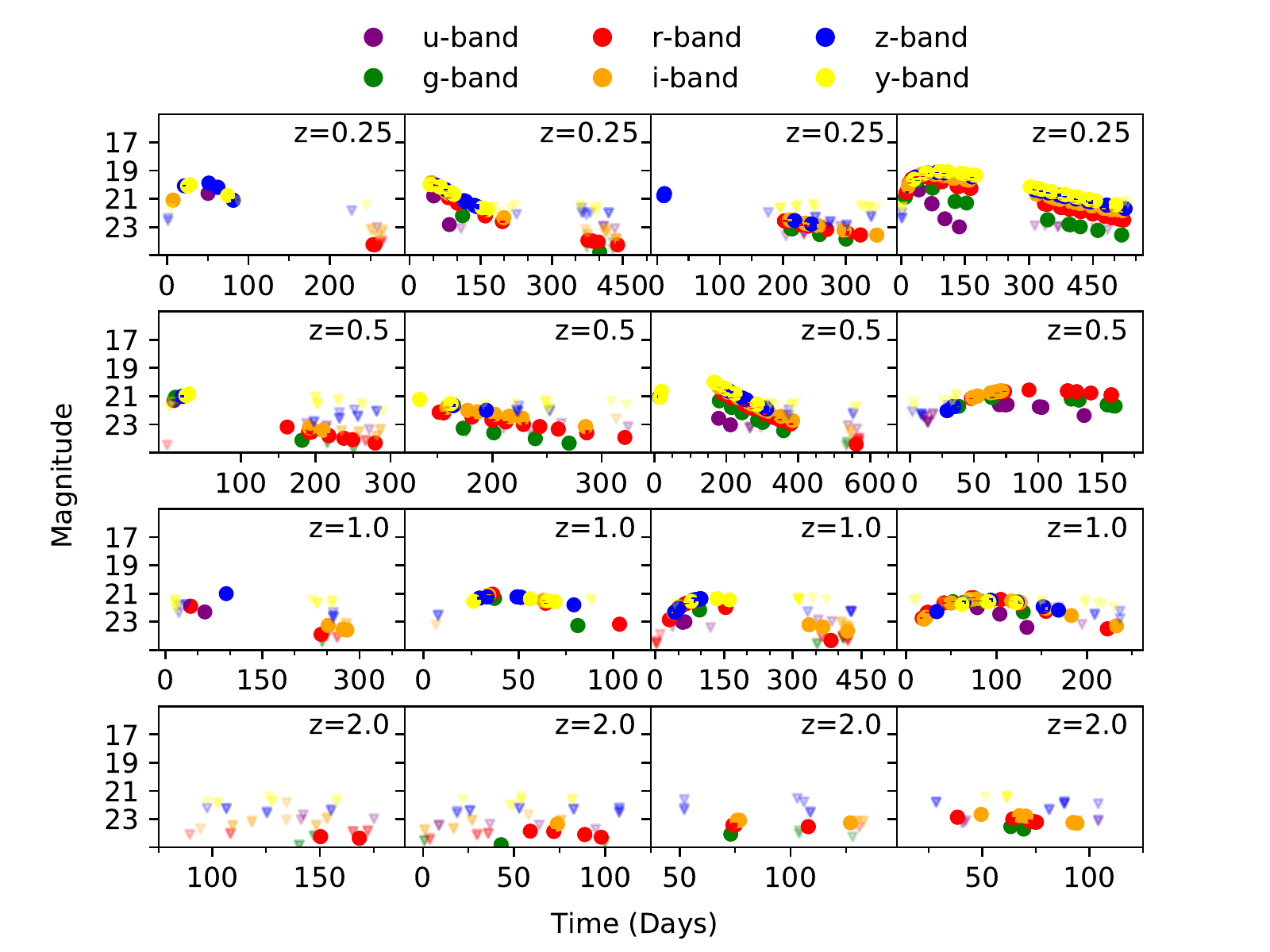}
\centering
\caption{Sample $ugrizy$ LSST light curves from {\tt OpSim} at four representative redshifts.  For each redshift, the light curves are ordered by the number of observations, with the left-most representing the bottom 10th percentile, the two middle panels representing the 50th percentile, and the right-most panel representing the 90th percentile.}
\label{fig:lcs_nofit}
\end{figure*}

\setlength\tabcolsep{2pt}
\label{tab:pars}
\begin{center}
\begin{deluxetable*}{cccccccccccccc} 
\tabletypesize{\footnotesize} 
\tablecolumns{14} 
\tablewidth{0pt} 
\tablecaption{SLSNe Parameters for Events Modeled in this Paper \label{table:results1}} 
\tablehead{ 
\colhead{} & \colhead{z} & \colhead{$P$} & \colhead{$B$} & \colhead{$M_\mathrm{ej}$} & \colhead{$v_\mathrm{ph}$} & \colhead{$E_\mathrm{min}$} & \colhead{$\kappa$} & \colhead{$\kappa_\gamma$} & \colhead{$M_\mathrm{NS}$}   & \colhead{$T_\mathrm{f}$}  &\colhead{$A_\mathrm{V}$}  &\colhead{$\sigma$}  &\colhead{WAIC$^*$}  \\
\colhead{}  & \colhead{} & \colhead{(ms)} & \colhead{$10^{14}$ G} & \colhead{M$_\odot$} & \colhead{$10^3$ km s\textsuperscript{-1}} & \colhead{$10^{51}$ erg} & \colhead{cm\textsuperscript{2}g\textsuperscript{-1}} & \colhead{cm\textsuperscript{2}g\textsuperscript{-1}}   & \colhead{M$_\odot$}  &\colhead{$10^3$ K}  &\colhead{mag}  &\colhead{mag}  &\colhead{}  \\\vspace{-0.2cm} 
}
\startdata 
PS1-11afv & 1.41 &  1.63$^{+ 0.57}_{- 0.68}$ &  0.20$^{+ 0.29}_{- 0.15}$ &  4.36$^{+ 4.75}_{- 1.85}$ & 11.21$^{+ 2.23}_{- 1.97}$ &  8.25$^{+ 7.52}_{- 3.87}$ &  0.14$^{+ 0.04}_{- 0.05}$ &  0.97$^{+10.94}_{- 0.92}$ &  1.87$^{+ 0.23}_{- 0.33}$ &  6.15$^{+ 1.10}_{- 1.13}$ &  0.26$^{+ 0.15}_{- 0.21}$ &  0.01$^{+ 0.03}_{- 0.01}$ & 34\\
PS1-11aib & 1.00 &  1.24$^{+ 0.28}_{- 0.22}$ &  0.60$^{+ 0.14}_{- 0.16}$ & 28.78$^{+10.89}_{- 5.68}$ &  5.44$^{+ 0.37}_{- 0.26}$ & 12.47$^{+ 5.91}_{- 2.16}$ &  0.11$^{+ 0.04}_{- 0.04}$ &  1.48$^{+36.52}_{- 1.43}$ &  1.83$^{+ 0.28}_{- 0.27}$ &  6.71$^{+ 1.02}_{- 1.27}$ &  0.44$^{+ 0.04}_{- 0.07}$ &  0.09$^{+ 0.02}_{- 0.01}$ & 137\\
PS1-11bdn & 0.52 &  3.47$^{+ 0.56}_{- 0.57}$ &  0.90$^{+ 0.30}_{- 0.23}$ &  1.09$^{+ 0.85}_{- 0.49}$ &  8.07$^{+ 1.17}_{- 1.27}$ &  1.10$^{+ 1.42}_{- 0.64}$ &  0.10$^{+ 0.06}_{- 0.04}$ & 21.65$^{+36.82}_{-14.03}$ &  1.82$^{+ 0.27}_{- 0.28}$ &  6.77$^{+ 0.82}_{- 0.95}$ &  0.13$^{+ 0.16}_{- 0.07}$ &  0.07$^{+ 0.05}_{- 0.03}$ & 15\\
PS1-11tt & 1.28 &  1.11$^{+ 0.34}_{- 0.25}$ &  0.06$^{+ 0.05}_{- 0.03}$ &  9.24$^{+ 4.30}_{- 3.27}$ & 13.77$^{+ 1.87}_{- 1.14}$ & 28.70$^{+11.37}_{-10.82}$ &  0.16$^{+ 0.03}_{- 0.04}$ &  0.03$^{+ 0.02}_{- 0.02}$ &  1.85$^{+ 0.23}_{- 0.30}$ &  6.07$^{+ 0.87}_{- 1.14}$ &  0.20$^{+ 0.18}_{- 0.14}$ &  0.01$^{+ 0.04}_{- 0.01}$ & 80\\
PS1-12bmy & 1.57 &  2.31$^{+ 0.29}_{- 0.37}$ &  0.52$^{+ 0.16}_{- 0.17}$ &  2.93$^{+ 1.17}_{- 0.70}$ & 13.10$^{+ 1.40}_{- 1.67}$ &  7.79$^{+ 2.33}_{- 1.71}$ &  0.14$^{+ 0.05}_{- 0.05}$ &  1.05$^{+14.40}_{- 0.95}$ &  1.90$^{+ 0.22}_{- 0.29}$ &  5.87$^{+ 1.05}_{- 1.10}$ &  0.36$^{+ 0.09}_{- 0.19}$ &  0.01$^{+ 0.04}_{- 0.00}$ & 47\\
PS1-12bqf & 0.52 &  5.15$^{+ 0.61}_{- 0.51}$ &  0.83$^{+ 0.24}_{- 0.16}$ &  2.82$^{+ 1.42}_{- 1.19}$ &  5.98$^{+ 0.46}_{- 0.44}$ &  1.51$^{+ 0.61}_{- 0.67}$ &  0.09$^{+ 0.06}_{- 0.03}$ &  4.21$^{+19.42}_{- 3.95}$ &  1.81$^{+ 0.29}_{- 0.28}$ &  6.06$^{+ 0.80}_{- 1.05}$ &  0.19$^{+ 0.16}_{- 0.14}$ &  0.01$^{+ 0.02}_{- 0.01}$ & 108\\
PS1-13gt & 0.88 &  1.03$^{+ 0.45}_{- 0.20}$ &  0.04$^{+ 0.03}_{- 0.01}$ &  7.04$^{+ 2.51}_{- 2.30}$ & 17.90$^{+ 3.58}_{- 3.39}$ & 35.96$^{+26.63}_{-19.34}$ &  0.16$^{+ 0.03}_{- 0.04}$ &  0.02$^{+ 0.03}_{- 0.01}$ &  1.75$^{+ 0.36}_{- 0.19}$ &  6.79$^{+ 0.76}_{- 0.67}$ &  0.27$^{+ 0.16}_{- 0.14}$ &  0.01$^{+ 0.03}_{- 0.01}$ & 76\\
PS1-13or & 1.52 &  1.82$^{+ 0.33}_{- 0.44}$ &  0.56$^{+ 0.31}_{- 0.25}$ &  8.43$^{+ 4.60}_{- 3.87}$ &  8.84$^{+ 1.16}_{- 0.69}$ &  9.37$^{+11.94}_{- 4.93}$ &  0.14$^{+ 0.04}_{- 0.05}$ &  0.48$^{+17.72}_{- 0.42}$ &  1.78$^{+ 0.30}_{- 0.27}$ &  5.99$^{+ 0.81}_{- 0.97}$ &  0.02$^{+ 0.04}_{- 0.01}$ &  0.10$^{+ 0.02}_{- 0.02}$ & 69\\
PTF10aagc & 0.21 &  4.48$^{+ 1.01}_{- 1.32}$ &  7.81$^{+ 1.35}_{- 1.74}$ &  1.62$^{+ 0.37}_{- 0.70}$ &  8.16$^{+ 0.49}_{- 0.44}$ &  1.63$^{+ 0.35}_{- 0.70}$ &  0.07$^{+ 0.05}_{- 0.01}$ &  7.34$^{+38.67}_{- 6.71}$ &  1.84$^{+ 0.20}_{- 0.27}$ &  5.98$^{+ 0.64}_{- 0.53}$ &  0.14$^{+ 0.16}_{- 0.11}$ &  0.19$^{+ 0.04}_{- 0.03}$ & 50\\
PTF10bfz & 0.17 &  1.21$^{+ 0.51}_{- 0.33}$ &  4.48$^{+ 0.95}_{- 1.02}$ & 10.23$^{+ 7.23}_{- 2.98}$ &  9.84$^{+ 1.03}_{- 0.73}$ & 17.29$^{+10.35}_{- 7.13}$ &  0.13$^{+ 0.05}_{- 0.05}$ &  0.88$^{+13.69}_{- 0.85}$ &  1.88$^{+ 0.25}_{- 0.28}$ &  5.23$^{+ 0.38}_{- 0.38}$ &  0.20$^{+ 0.16}_{- 0.14}$ &  0.09$^{+ 0.02}_{- 0.02}$ & 82\\
PTF10nmn & 0.12 &  2.21$^{+ 1.46}_{- 0.92}$ &  1.03$^{+ 0.33}_{- 0.27}$ &  5.12$^{+ 3.64}_{- 2.54}$ & 10.38$^{+ 3.40}_{- 3.37}$ &  7.06$^{+ 9.53}_{- 3.72}$ &  0.13$^{+ 0.05}_{- 0.05}$ &  0.43$^{+ 0.47}_{- 0.27}$ &  1.87$^{+ 0.23}_{- 0.31}$ &  7.22$^{+ 0.47}_{- 0.36}$ &  0.19$^{+ 0.17}_{- 0.11}$ &  0.13$^{+ 0.05}_{- 0.06}$ & 35\\
PTF10uhf & 0.29 &  4.35$^{+ 1.52}_{- 2.47}$ &  3.19$^{+ 1.72}_{- 1.12}$ &  2.39$^{+ 3.14}_{- 0.79}$ & 10.54$^{+ 1.10}_{- 1.49}$ &  3.69$^{+ 6.20}_{- 1.51}$ &  0.15$^{+ 0.04}_{- 0.04}$ &  1.56$^{+ 9.36}_{- 1.49}$ &  1.84$^{+ 0.25}_{- 0.30}$ &  5.94$^{+ 1.17}_{- 1.03}$ &  0.35$^{+ 0.10}_{- 0.21}$ &  0.14$^{+ 0.04}_{- 0.03}$ & 27\\
PTF10vqv & 0.45 &  3.12$^{+ 0.42}_{- 0.68}$ &  1.43$^{+ 0.29}_{- 0.34}$ &  2.71$^{+ 2.51}_{- 1.03}$ &  9.30$^{+ 1.61}_{- 1.80}$ &  3.59$^{+ 1.75}_{- 1.31}$ &  0.11$^{+ 0.06}_{- 0.04}$ &  3.44$^{+34.25}_{- 2.92}$ &  1.87$^{+ 0.23}_{- 0.26}$ &  8.72$^{+ 0.63}_{- 0.73}$ &  0.05$^{+ 0.07}_{- 0.03}$ &  0.01$^{+ 0.02}_{- 0.01}$ & 35\\
PTF11hrq & 0.06 &  7.68$^{+ 1.37}_{- 0.90}$ &  1.15$^{+ 0.29}_{- 0.28}$ &  1.73$^{+ 0.34}_{- 0.44}$ & 11.36$^{+ 1.50}_{- 2.08}$ &  3.30$^{+ 0.99}_{- 1.03}$ &  0.16$^{+ 0.03}_{- 0.04}$ &  0.38$^{+ 0.14}_{- 0.14}$ &  1.89$^{+ 0.22}_{- 0.30}$ &  5.31$^{+ 1.40}_{- 0.33}$ &  0.19$^{+ 0.17}_{- 0.12}$ &  0.06$^{+ 0.01}_{- 0.01}$ & 170\\
PTF12gty & 0.18 &  6.38$^{+ 0.61}_{- 0.97}$ &  1.56$^{+ 0.34}_{- 0.32}$ &  7.92$^{+ 1.53}_{- 1.43}$ &  5.67$^{+ 0.44}_{- 0.21}$ &  3.95$^{+ 0.94}_{- 0.83}$ &  0.17$^{+ 0.02}_{- 0.04}$ &  1.85$^{+30.95}_{- 1.69}$ &  1.84$^{+ 0.25}_{- 0.28}$ &  5.71$^{+ 0.34}_{- 0.33}$ &  0.27$^{+ 0.16}_{- 0.17}$ &  0.13$^{+ 0.03}_{- 0.02}$ & 97\\
PTF12hni & 0.11 &  6.85$^{+ 2.00}_{- 1.82}$ &  3.21$^{+ 0.67}_{- 0.67}$ &  3.77$^{+ 1.93}_{- 1.73}$ &  6.67$^{+ 1.12}_{- 1.09}$ &  2.56$^{+ 0.90}_{- 0.89}$ &  0.13$^{+ 0.04}_{- 0.05}$ &  3.23$^{+25.48}_{- 2.72}$ &  1.84$^{+ 0.23}_{- 0.26}$ &  5.22$^{+ 0.34}_{- 0.25}$ &  0.15$^{+ 0.20}_{- 0.10}$ &  0.33$^{+ 0.03}_{- 0.03}$ & 39\\
PTF12mxx & 0.33 &  2.24$^{+ 0.44}_{- 0.48}$ &  0.95$^{+ 0.35}_{- 0.27}$ &  6.71$^{+ 2.36}_{- 1.76}$ &  8.16$^{+ 0.51}_{- 0.35}$ &  6.78$^{+ 2.83}_{- 2.12}$ &  0.16$^{+ 0.03}_{- 0.05}$ &  0.03$^{+ 4.70}_{- 0.02}$ &  1.85$^{+ 0.22}_{- 0.32}$ &  6.04$^{+ 0.89}_{- 0.88}$ &  0.26$^{+ 0.11}_{- 0.13}$ &  0.05$^{+ 0.01}_{- 0.01}$ & 197\\
PTF13bdl & 0.40 &  1.09$^{+ 0.34}_{- 0.28}$ &  1.40$^{+ 0.46}_{- 0.33}$ & 68.80$^{+19.94}_{-21.18}$ &  5.76$^{+ 0.78}_{- 0.48}$ & 35.18$^{+17.68}_{-12.45}$ &  0.18$^{+ 0.02}_{- 0.03}$ &  0.90$^{+25.42}_{- 0.86}$ &  1.84$^{+ 0.26}_{- 0.31}$ &  5.97$^{+ 0.80}_{- 1.11}$ &  0.36$^{+ 0.08}_{- 0.17}$ &  0.01$^{+ 0.02}_{- 0.01}$ & 59\\
PTF13bjz & 0.27 &  2.98$^{+ 2.19}_{- 1.67}$ &  5.51$^{+ 2.87}_{- 2.91}$ &  2.07$^{+ 2.26}_{- 0.68}$ & 11.09$^{+ 2.05}_{- 2.22}$ &  3.90$^{+ 5.66}_{- 2.20}$ &  0.13$^{+ 0.05}_{- 0.05}$ &  0.66$^{+14.75}_{- 0.62}$ &  1.75$^{+ 0.31}_{- 0.22}$ &  5.88$^{+ 1.01}_{- 0.81}$ &  0.21$^{+ 0.19}_{- 0.14}$ &  0.18$^{+ 0.04}_{- 0.03}$ & 19\\
PTF13cjq & 0.40 &  1.75$^{+ 0.85}_{- 0.54}$ &  1.49$^{+ 0.31}_{- 0.38}$ &  7.75$^{+ 4.52}_{- 2.93}$ &  8.33$^{+ 1.26}_{- 1.13}$ &  6.96$^{+ 4.55}_{- 1.76}$ &  0.12$^{+ 0.04}_{- 0.04}$ & 10.02$^{+44.17}_{- 8.14}$ &  1.86$^{+ 0.23}_{- 0.33}$ &  6.83$^{+ 0.66}_{- 0.52}$ &  0.10$^{+ 0.11}_{- 0.07}$ &  0.22$^{+ 0.03}_{- 0.02}$ & 63
\enddata 
\tablecomments{$^*$The Watanabe-Akaike information criteria (or ``widely applicable Bayesian criteria"; \citealt{watanabe2010asymptotic,gelman2014understanding}} 
\end{deluxetable*}
\end{center}

\subsection{Description of the LSST Simulation}

After generating the sample of SLSN models, we inject the simulated events into {\tt OpSim}, a publicly available application that simulates LSST's scheduler and image acquisition process over its 10-year survey. {\tt OpSim} realistically accounts for the science program requirements, mechanics of the telescope design and potential environmental conditions to produce a database of observations. We use {\tt OpSim} to calculate the estimated signal-to-noise ratios and limiting magnitudes of each observation, using the formulae outlined in the Appendix. {\tt OpSim} offers a number of unique schedulers, each designed to optimize distinct scientific goals; we use the most recent simulation, dubbed {\tt minion\_1016} \citep{delgado2014lsst}.

For both the WFD survey and DDFs, we inject our simulated models uniformly at $z=0-6$ in bins of $\Delta z=0.1$. In each bin, a sample of 1,000 models are randomly injected uniformly across the sky and in time to calculate the ``discovery efficiency'' (see \S\ref{sec:eff}) of the LSST observing strategy as a function of redshift.  We resample the simulated models to the observed times and add white noise corresponding to the estimated signal-to-noise ratio reported by {\tt OpSim}. Additionally, we add Milky Way extinction based on the injected sky positions. We disregard host galaxy extinction since most known SLSN host galaxies appear to have negligible extinction (e.g.,  \citealt{chen2015host,leloudas2015spectroscopy,nicholl2017magnetar,lunnan2014hydrogen}). Example light curves at representative redshifts are shown in Figure~\ref{fig:lcs_nofit}; featured light curves are selected to highlight a combination of best, typical and worst cases in the WFD survey. The luminosity-duration phase space of our simulated SLSNe are shown in Figure \ref{fig:dlps}.

\section{Characteristics of SLSNe discovered by LSST} \label{sec:char}

The thousands of injected simulated light curves reflect the wide range of observed SLSN properties expected from LSST.  Here we summarize these properties, define our criteria for detection, and determine the rate of detected SLSNe as a function of redshift for both the WFD survey and DDFs. 

We calculate the expected number of SLSNe within each redshift bin by multiplying the sample recovered from {\tt OpSim} by the estimated volumetric rate from \citet{quimby2013rates} normalized to the cosmic star formation history \citep{madau2014}: 
\begin{equation}
\label{eq:rate_eq}
R = R_0\frac{(1+z)^{2.7}}{1+[(1+z)/2.9]^{5.6}} \,	\mathrm{Gpc}^{-3}\mathrm{yr}^{-1},
\end{equation}
where $R_0\approx 21\,\mathrm{Gpc}^{-3}\mathrm{yr}^{-1}$ is the normalized SLSNe rate at $z=0$ with an uncertainty range of $R_0\approx 4-72$ Gpc$^{-3}$ \citep{quimby2013rates,prajs2016volumetric}. Using this prescription, the volumetric SLSN rate peaks at $z\approx 1.5-2$.

\begin{figure*}[t!]
\includegraphics[width=0.95\textwidth]{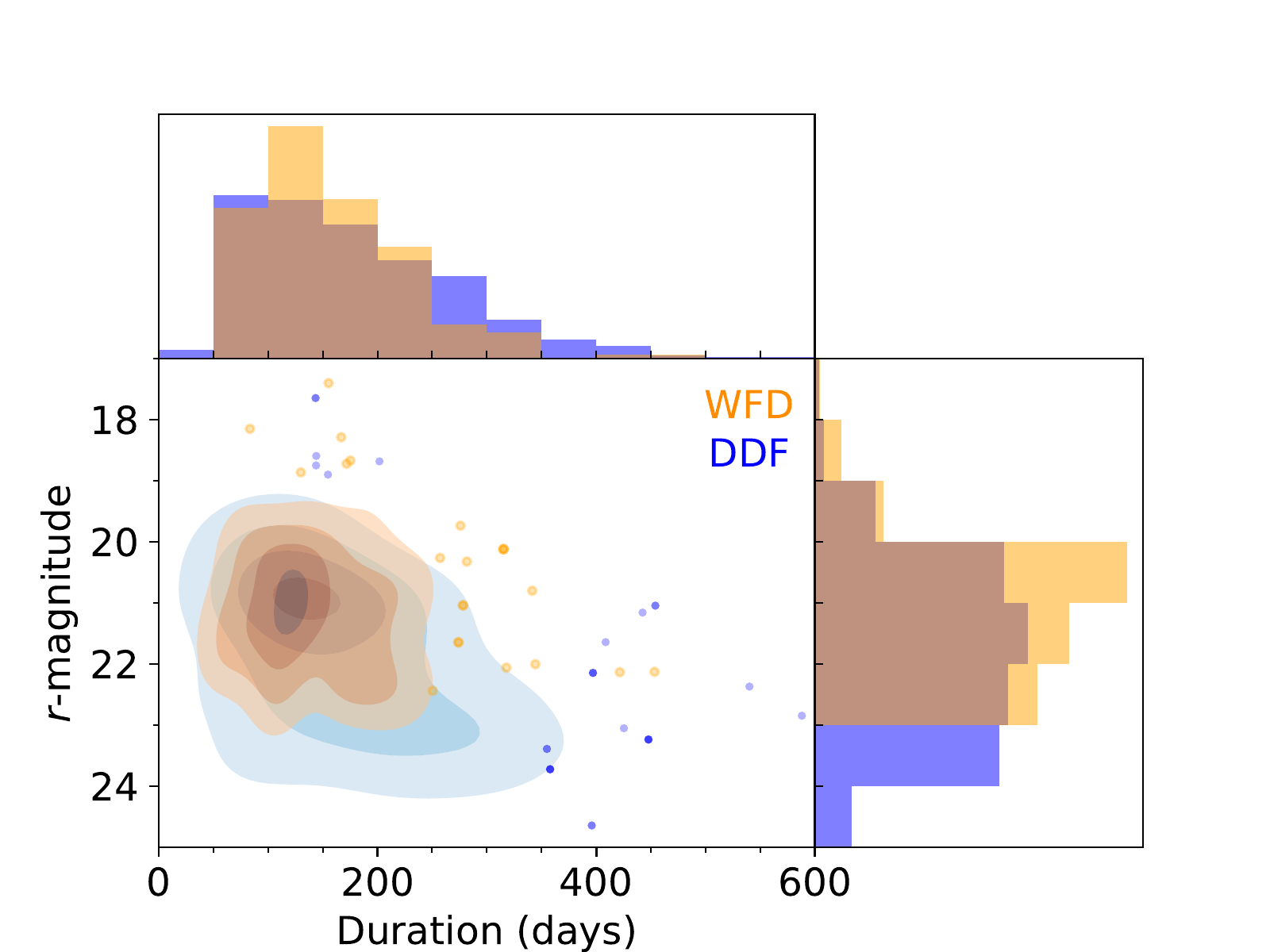}
\centering
\caption{Duration-luminosity phase space for a representative sample of our simulated SLSNe as observed with LSST in the WFD survey (orange) and DDFs (blue). Contours are a kernel density estimate of the two populations, while points represent outliers. The majority of objects have a peak magnitude of $r\approx 21-22$ mag and durations of $\approx 100$ days.  The events in the DDFs extend to lower peak magnitudes and longer durations.}
\label{fig:dlps}
\end{figure*}

We first focus on the $r$-band light curves to provide an overview of the broad observational properties. In Figure~\ref{fig:dlps}, we show the observed duration-luminosity phase space for our injected light curves (weighted by their volumetric rate) using a kernel density estimate. For the WFD survey, peak observed magnitudes span $\approx 19-23$ mag, with the distribution peaking at $\approx 21.2$ mag.  We also find that the observed durations of the SLSNe span $t_\mathrm{dur}\approx 60-300$ days. This timescale is comparable to the expected LSST season length ($\approx 4-6$ months), implying that for a substantial fraction of events the rise or decline will be missed in seasonal gaps.  We find that a typical SLSN is tracked for $\approx 100-400$ days ($1\sigma$ uncertainty range). For the DDFs, we find that average peak magnitudes are slightly dimmer, with the distribution peaking at $\approx 21.8$ mag. The observed durations are similar to those in the WFD survey.

\begin{figure}[t!]
\includegraphics[width=0.48\textwidth]{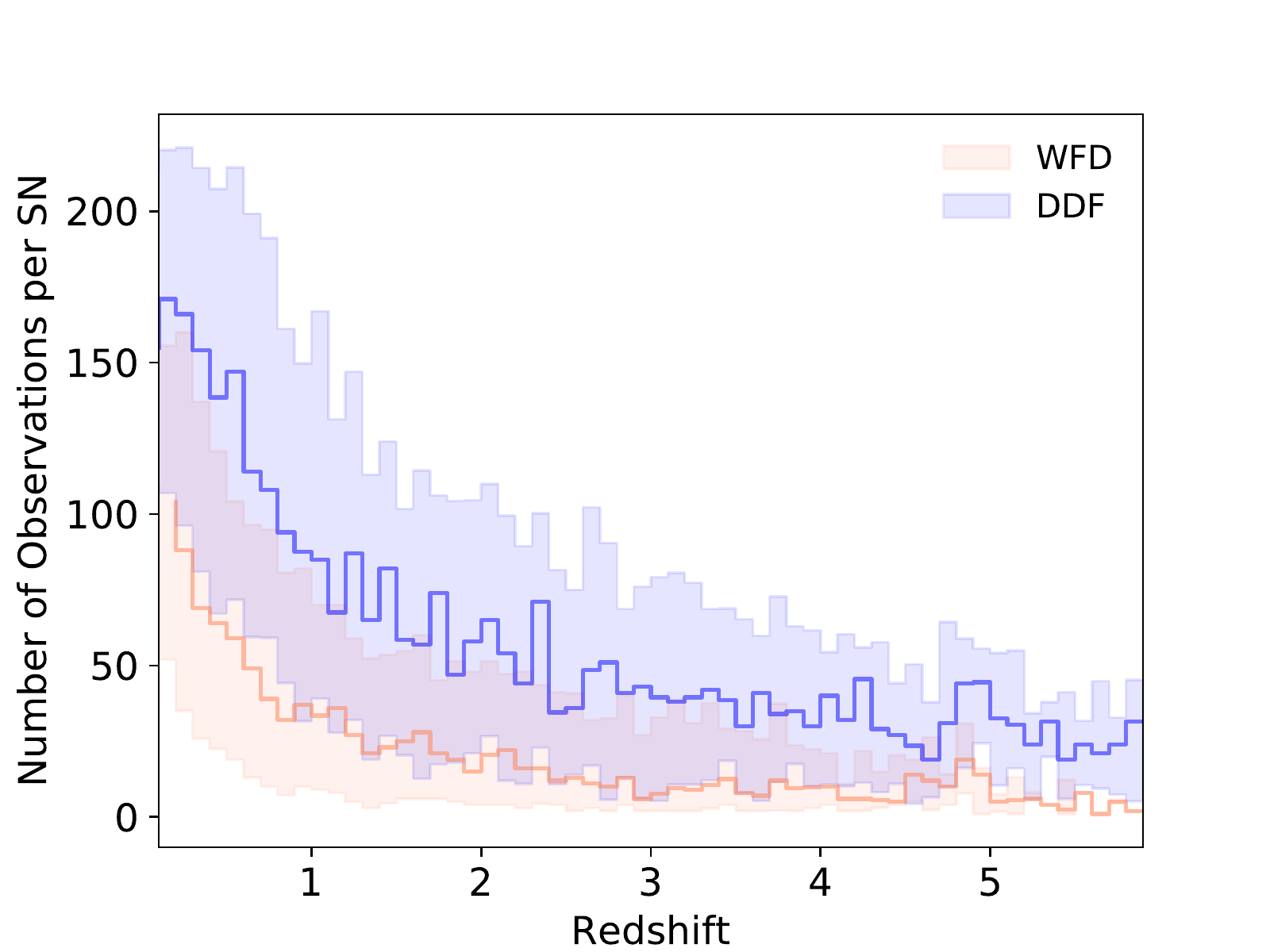}
\centering
\caption{Number of observed data points per event as a function of redshift. Bold lines show the mean in each redshift bin, with the shaded regions representing $1\sigma$ ranges due to event-to-event variations. The number of observed points drops to $<10$ (our minimum criterion for a detection) at $z\approx3$ in the WFD survey, while it typically remains at $>20$ in the DDFs even to $z\approx 6$.}
\label{fig:num_vs_z}
\end{figure}

The total number of observed light curve data points (combined in all filters) rapidly decreases with redshift; see Figure~\ref{fig:num_vs_z}.  For the WFD survey, SLSNe at $z\lesssim 1$ will have $\approx 50-100$ data points, while the majority of SLSNe have light curves with $\lesssim 50$ data points. The number of observed data points roughly doubles for the SLSNe in the DDFs due to both their higher cadence and deeper limits, enabling a longer temporal baseline.

\begin{deluxetable*}{c|cccc|cccc} 
\tablecolumns{9} 
\tablewidth{0pt} 
\tablecaption{Observational Metrics \label{table:metrics}} 
\tablehead{ 
\colhead{}& \multicolumn{4}{c}{WFD}  & \multicolumn{4}{c}{DDF} \\
\colhead{Metric} & 
\colhead{Limiting $z^1$} & 
\colhead{Discovered$^2$} & 
\colhead{Useful} & 
\colhead{Useful} & 
\colhead{Limiting z$^1$}& 
\colhead{Discovered$^2$} & 
\colhead{Useful} & 
\colhead{Useful}  \\
\colhead{}& \colhead{}& 
\colhead{(yr$^{-1}$)} & 
\colhead{(lenient)$^3$} & 
\colhead{(strict)$^4$} & 
\colhead{}& 
\colhead{(yr$^{-1}$)} & 
\colhead{(lenient)$^3$} & 
\colhead{(strict)$^4$}
}
\startdata 
$>10$ detections  &  2.9  &  9600  &  4810 (0.50)  &  1680 (0.18)  & 3.9  &  13  &  5 (0.40)  &  1 (0.13) \\
$>50$ detections  &  2.3  &  3320  &  2170 (0.65)  &  900 (0.27)  & 3.5  &  9  &  4 (0.50)  &  1 (0.17) \\
$>100$ detections  &  2.0  &  590  &  416 (0.71)  &  216 (0.37)  & 2.7  &  3  &  2 (0.62)  &  0 (0.24) \\
$>10$ detections in $u$  &  2.5  &  50  &  30 (0.65)  &  20 (0.37)  & 2.4  &  4  &  2 (0.61)  &  0 (0.22) \\
$>10$ detections in $g$  &  2.3  &  1090  &  720 (0.66)  &  320 (0.29)  & 3.1  &  6  &  3 (0.55)  &  1 (0.20) \\
$>10$ detections in $r$  &  2.8  &  6600  &  3190 (0.48)  &  1170 (0.18)  & 3.7  &  10  &  4 (0.40)  &  1 (0.13) \\
$>10$ detections in $i$  &  2.7  &  6120  &  3170 (0.52)  &  1170 (0.19)  & 3.8  &  10  &  4 (0.41)  &  1 (0.13) \\
$>10$ detections in $z$  &  2.3  &  2770  &  1730 (0.63)  &  730 (0.27)  & 3.8  &  10  &  4 (0.45)  &  1 (0.15) \\
$>10$ detections in $y$  &  2.0  &  1300  &  890 (0.69)  &  420 (0.33)  & 3.2  &  6  &  3 (0.53)  &  1 (0.19) \\
$>10$ observations in rise  &  3.6  &  2240  &  577 (0.26)  &  131 (0.06)  & 4.0  &  8  &  2 (0.35)  &  0 (0.10) \\
$>20$ during peak$^5$  &  2.2  &  2690  &  1740 (0.65)  &  780 (0.29)  & 3.4  &  9  &  4 (0.49)  &  1 (0.17) \\
Detected $<5$ days after explosion  &  2.9  &  100  &  36 (0.36)  &  13 (0.14)  & 3.7  &  2  &  0 (0.31)  &  0 (0.10) \\
Detected $<10$ days after explosion  &  3.2  &  350  &  119 (0.34)  &  42 (0.12)  & 3.9  &  5  &  1 (0.28)  &  0 (0.09) \\
Measurable duration in $u$  &  1.2  &  50  &  39 (0.77)  &  23 (0.46)  & 1.8  &  0  &  0 (0.71)  &  0 (0.34) \\
Measurable duration in $g$  &  2.1  &  280  &  185 (0.66)  &  78 (0.28)  & 2.2  &  1  &  0 (0.65)  &  0 (0.27) \\
Measurable duration in $r$  &  2.2  &  960  &  618 (0.65)  &  261 (0.27)  & 2.7  &  2  &  1 (0.59)  &  0 (0.22) \\
Measurable duration in $i$  &  2.0  &  770  &  514 (0.67)  &  221 (0.29)  & 3.1  &  2  &  1 (0.60)  &  0 (0.22) \\
Measurable duration in $z$  &  2.1  &  310  &  215 (0.70)  &  108 (0.35)  & 3.2  &  2  &  1 (0.57)  &  0 (0.22) \\
Measurable duration in $y$  &  2.0  &  90  &  64 (0.69)  &  41 (0.44)  & 2.0  &  0  &  0 (0.63)  &  0 (0.26)
\enddata 
\tablecomments{$^1$Cumulative redshift at which 90\% of SLSNe are discovered.\\
$^2$Total number of discovered SLSNe satisfying given metric. \\$^3$Total Number of discovered SLSNe satisfying given metric with recoverable parameters to within a factor of two.\\
$^4$Total Number of discovered SLSNe satisfying given metric with recoverable parameters to within 30\%.\\
$^5$Here we define ``peak'' as within one magnitude of peak brightness.} 
\end{deluxetable*}

\subsection{Efficiency and Metrics for Detectability}
\label{sec:eff}

We now focus on more quantitative measures of ``detectability" for the simulated light curves. There can be many definitions of a detection of a transient. We are most interested in the ability to: (1) accurately estimate the physical parameters from the light curve, and (2) discover events sufficiently early to enable follow up with other instruments. To address these points, we quantify the information content of the observed light curves by defining several properties that can be easily measured directly from the light curve. We focus on 19 representative properties, summarized in Table~\ref{table:metrics}. 

\begin{figure}[t!]
\includegraphics[width=0.48\textwidth]{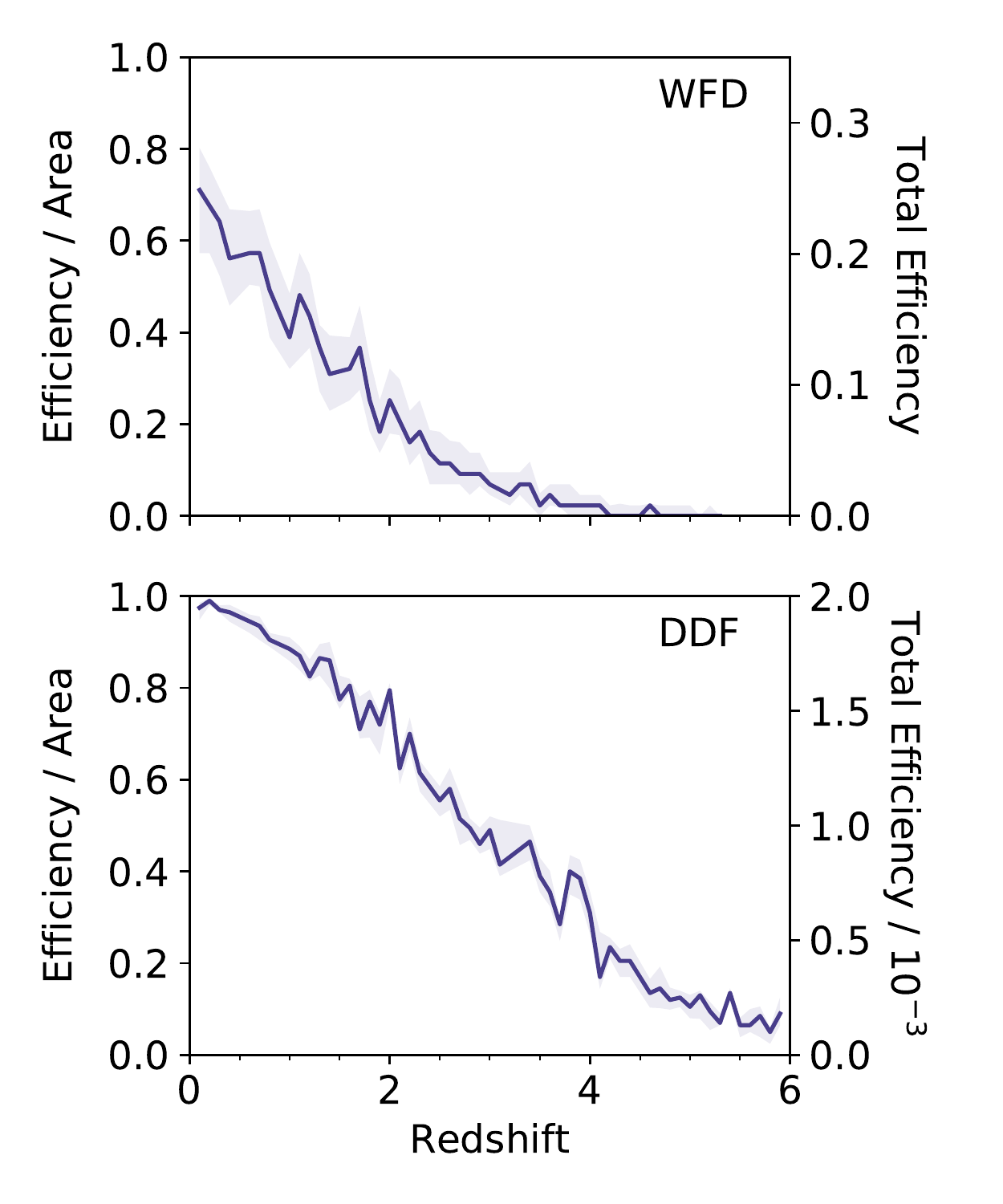}
\centering
\caption{WFD survey and DDFs efficiencies as a function of redshift. The left-hand y-axis shows the efficiency of SLSN detection, assuming the SLSNe are within the survey footprint (i.e., a SLSN may be in the footprint but too dim to detect). The right-hand y-axis shows the total efficiency, $\epsilon$, including the effect of survey area. The shaded region represents the $1\sigma$ error bars from bootstrap analysis.}
\label{fig:eff}
\end{figure}

Using each property as an independent criterion of detection, we calculate the total detection efficiency for SLSNe in LSST. The efficiency can be divided into two multiplicative parts. The first is the survey efficiency, $\epsilon_s$, arising from the survey footprint and cadence. We calculate this by injecting events uniformly across the complete survey duration and the sky.  We then calculate the fraction of events that are in the LSST footprint during at least one observation. This efficiency is effectively the area covered by the survey, given the long duration of SLSNe compared to the cadence of LSST.  The second efficiency, $\epsilon_m$, is the fraction of simulated events that satisfy the metrics listed in Table~\ref{table:metrics}. 

The total efficiency, $\epsilon \equiv \epsilon_s \times \epsilon_m$, as a function of redshift is shown in Figure~\ref{fig:eff} for light curves with $>10$ observations (one of our 19 metrics). For both the WFD survey and DDFs, the efficiencies decline monotonically as a function of redshift; however, the decline is shallower in the case of the deeper DDFs. Within the WFD survey footprint, the efficiency peaks at $\approx 70$\% at low redshift; in the DDF fields, the efficiency peaks at $\approx 100$\%. The WFD survey peaks at a lower efficiency due to the fact that SLSNe (particularly if they explode outside of the observing season for their part of the sky) can be discovered well beyond peak, at which point they may already be below the LSST detection limit. The WFD survey efficiency reaches $50\%$ at $z\approx 1$ and declines to $10\%$ at $z\approx 3$. For the DDFs, the efficiency reaches $50\%$ at $z\approx 3$ and declines to $10\%$ at $z\approx 5$. 

\begin{figure*}[t]
\includegraphics[width=0.95\textwidth]{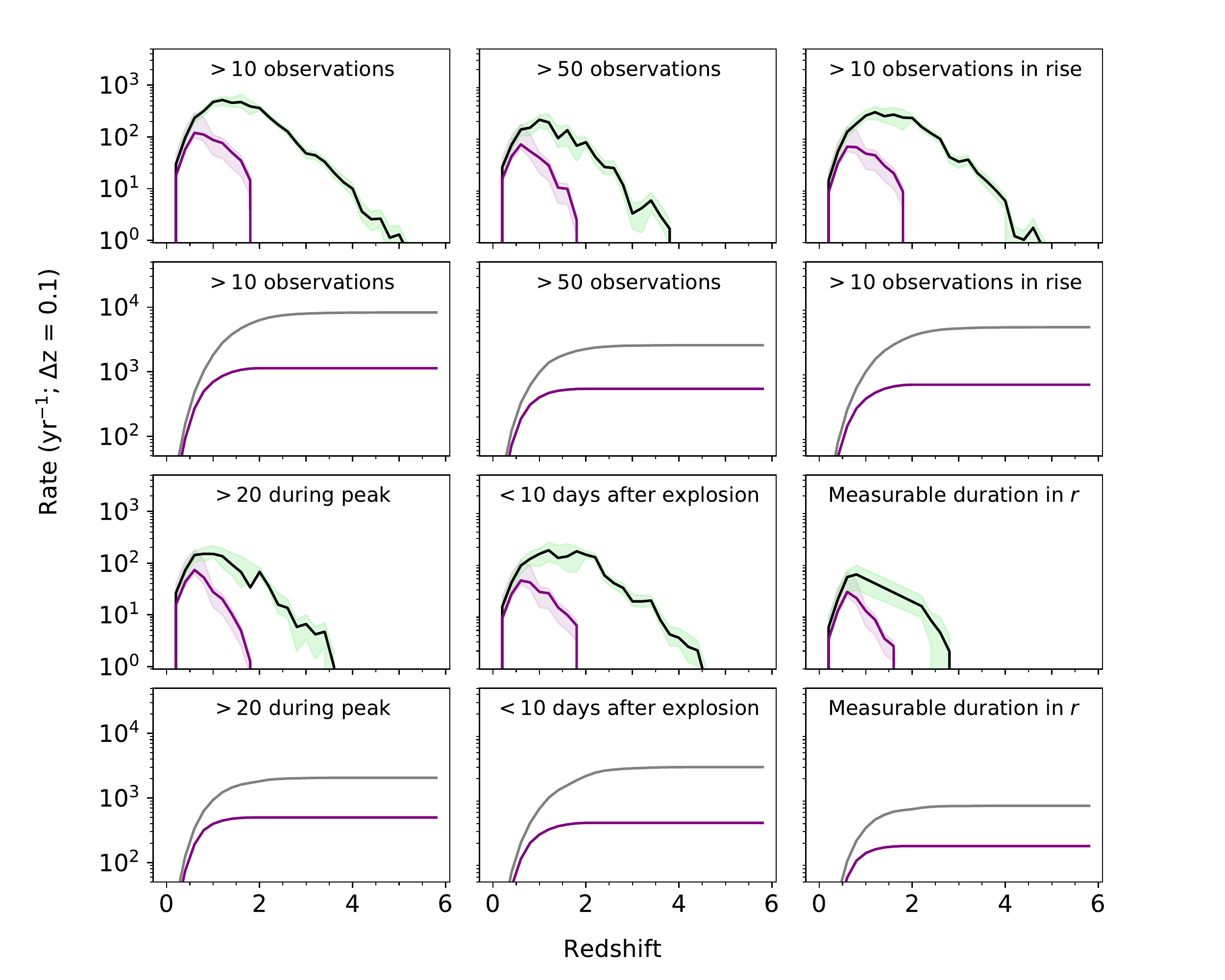}
\centering 
\caption{{\it First and third rows}: The WFD survey annual detection rate of SLSNe as a function of redshift (black lines) for various metrics.  The green shaded regions represent  $1\sigma$ errors from our bootstrap analysis. Also shown are the rates for SLSNe with (strict) recoverable parameters (purple line and shaded area); note that the purple line is calculated assuming the same information efficiency for each metric. {\it Second and fourth rows}: Cumulative distributions of SLSNe that satisfy each metric (black) and those that have lenient recoverable parameters (purple).}
\label{fig:wfd_num}
\end{figure*}

\begin{figure*}[t]
\includegraphics[width=0.95\textwidth]{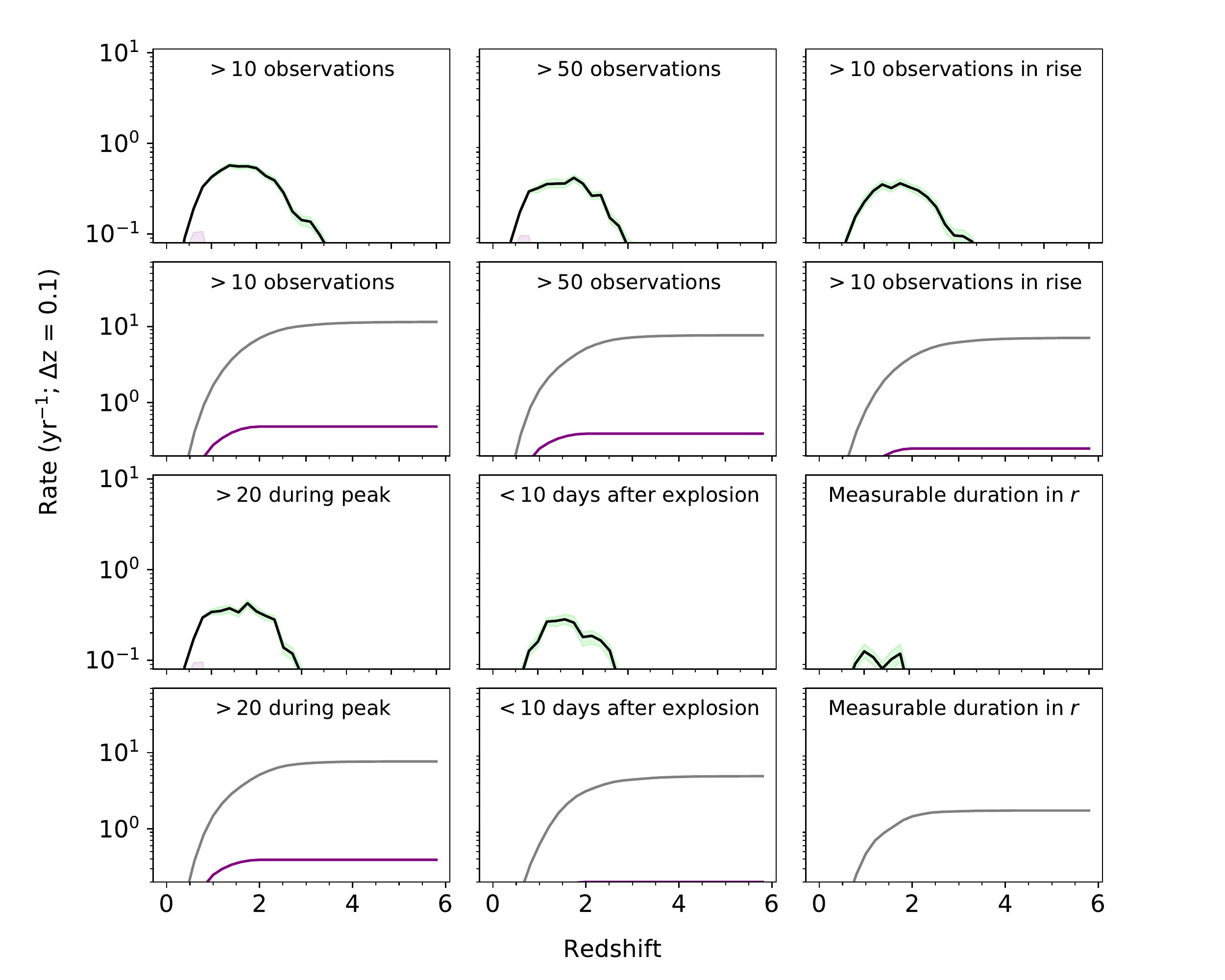}
\centering 
\caption{Same as Figure~\ref{fig:wfd_num} but for each DDF.}
\label{fig:ddf_num}
\end{figure*}

We combine the efficiency and the estimated volumetric rate to calculate the total expected number of SLSNe, the integral of the observed rate over the comoving volume, corrected for time dilation:
\begin{equation}
\label{eq:num_eq}
N = \epsilon\int_{z_\mathrm{min}}^{z_\mathrm{max}} \frac{4\pi R}{1+z} \frac{dV}{dz} dz.
\end{equation}
We estimate the statistical uncertainty in the number of detections due to the uncertainty in $\epsilon$ using a bootstrap analysis; namely, we resample the properties of the observed light curves repeatedly, recalculating the efficiencies each time. There is an overall scaling uncertainty due to the systematic uncertainty in the volumetric rate, but we expect this to be improved with new rate measurements from DES \citep{des2005} and ZTF \citep{ztfcitation}.  Therefore, in Figures~\ref{fig:eff}--\ref{fig:ddf_num} we show only the statistical uncertainties.

The number of SLSNe discovered per year as a function of redshift is shown in Figures~\ref{fig:wfd_num} and \ref{fig:ddf_num}. For our most lenient definition of a detection (at least 10 light curve points), we find that LSST will discover $\approx 9,600$ SLSNe per year. This is in agreement with the rate reported in the Scientific Handbook \citep{lsstsciencecollab} and previous studies \citep{tanaka2013detectability,scovacricchi2015cosmology}. The distribution roughly traces the cosmic star formation history to $z\approx 1$, at which point the observed distribution drops more rapidly due to the declining detection efficiency. The distribution extends to $z\approx 3$ in the WFD survey and to $z\approx 5$ in the DDFs.

Measuring both the peak brightness and duration can lead to robust measurements of the SLSN properties. About 2,700 SLSNe per year will have more than 20 observations within one magnitude of their peak brightness. We find that the duration is most readily measurable in $r$-band, and that for about 950 SLSNe per year $t_\mathrm{dur}$ (as defined above) will be measured; we note that this represents only 10\% of the overall SLSN sample.

Capturing the light curve rise can be especially important for constraining the underlying power source in SLSNe and to search for early bumps in the light curve. We find that $\approx 800$ SLSNe per year will be discovered within ten days (in the rest frame) of their explosion time, comparable to the typical time frame of the early pre-peak bumps seen to date \citep{nicholl2016seeing,leloudas2012sn,nicholl2015lsq14bdq,smith2016des14x3taz}. About $100$ SLSNe per year will be discovered within five days of explosion, most being located at $z\lesssim 1$. About $4,200$ SLSNe will be detected with at least ten observations during the rising phase; however, fewer than 10 SLSNe with this property will be found annually in each DDF, due to their small areal footprint.

\section{Recovering the SLSN Parameters}
\label{sec:recover}

In the previous section we explored the overall detection rates and the redshift distributions for a range of observational light curve metrics.  Here we fit the simulated LSST light curves with the same model used to generate them to determine how well we can recover the injected model parameters.  Our goals are to understand how well we can determine the model parameters from LSST data, and to correlate our simple light curve metrics to the information content of the light curves. The latter goal is important because the final survey strategies of LSST will be determined by providing a simple, measurable metric that can be optimized for a specific science goal. 

\subsection{Injection and Recovery of Representative SLSNe}

We fit the output light curves using {\tt MOSFiT} in the same manner that was used to generate them (\S\ref{sec:sim}).  We focus on three SLSNe representative of the larger population: SN2013dg ($t_\mathrm{dur}\approx45$ days), LSQ12dlf ($t_\mathrm{dur}\approx60$ days), and SN2015bn ($t_\mathrm{dur}\approx130$ days).  All three have roughly the same peak luminosity, $\approx 3\times10^{44}$ erg s$^{-1}$. We inject and recover about 100 iterations of these SLSNe at  $z=0.5,1.0,1.5,2.0,3.0$. 

We fix the redshift to its input value when fitting, finding that without doing so it is nearly impossible to constrain the explosion parameters. In reality, it is unclear how well we will know the redshift a priori through photometry or (in some cases) spectroscopic measurements of their host galaxies. SLSNe are typically found in low-luminosity ($M_\mathrm{B}\approx-17$ mag) host galaxies \citep{lunnan2014}. For $z\lesssim 0.5$, most of these hosts will fall in the so-called LSST ``gold'' galaxy sample (defined as galaxies with $m_\mathrm{i}<25.3$ mag), which will have a root-mean-square scatter in the photometric redshifts of $\sigma_\mathrm{z}/(1+z)\lesssim 0.05$ (see \citealt{lsstsciencecollab}). We additionally fix the host reddening to be negligible.

We are interested in our ability to recover four key parameters: the ejecta mass, the ejecta velocity, the initial magnetar spin period, and the magnetic field. In our model, the spin period and magnetic field have strong degeneracies with several nuisance parameters, making them difficult to directly measure. We therefore recover the following variables, which directly correlate with the rotational energy and spindown timescales of the magnetar \citep{nicholl2017magnetar,villar2017theoretical}:
$$B^*\equiv B^{-2}(\sin{\theta})^{-2}M_\mathrm{NS}^{3/2}$$
$$P^*\equiv P_\mathrm{spin}^{-2}M_\mathrm{NS}^{3/2}$$,
where $\theta$ is the angle between the rotational axis and magnetic dipole, and $M_\mathrm{NS}$ is the neutron star mass.

How well we need to recover the SLSN parameters depends on the scientific goal. For cosmological studies, determining the average distance modulus (assuming SLSNe are standardizable; see \citealt{inserra2014superluminous}) to $\approx 0.25$ mag is sufficient to constrain, for example, $\Omega_m$ to within 2\% \citep{scovacricchi2015cosmology}.  In \citet{nicholl2017magnetar}, constraining parameters to an average of $\approx30-50$\% was sufficient to probe the underlying  population with a sample of $\approx 50$ events. We track our ability to recover the four key parameters to (1) $\lesssim 30$\% of their input values and with error bars of $<50$\% (``strict''), and (2) within a factor of two of their input values with error bars of $<50$\% (``lenient").

\begin{figure*}[t] 
\includegraphics[width=0.95\textwidth]{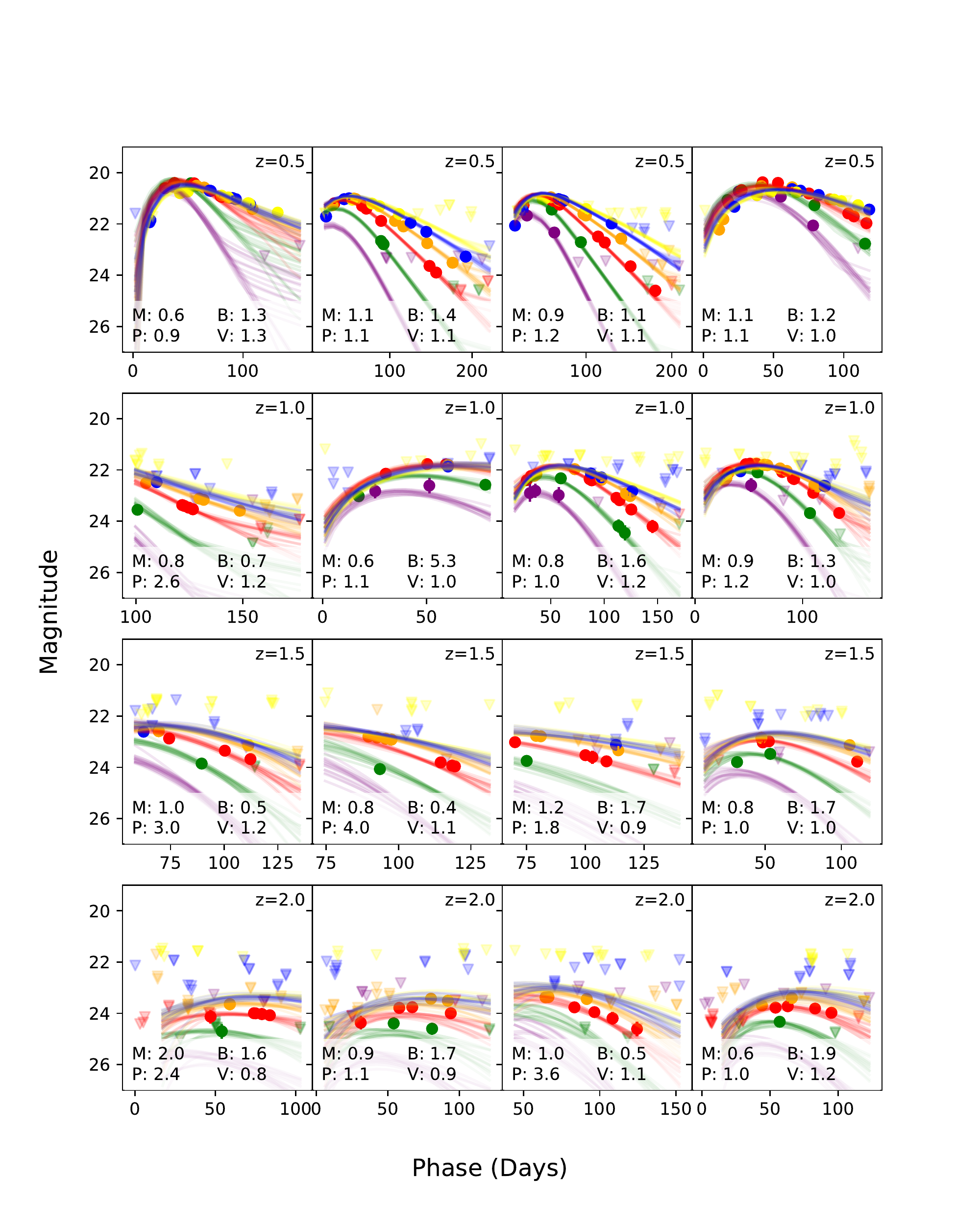}
\centering
\caption{Light curves fits from {\tt MOSFiT} to a sample of simulated LSST light curves at representative redshifts.  For each event we list the ratio of the injected to fitted values for the four key model parameters. The light curves are ordered by the quality of parameter recovery from left (worst) to right (best).}
\label{fig:simulated_lc}
\end{figure*}

Example light curves and their best-fit models are shown in Figure~\ref{fig:simulated_lc}. At low redshifts, many of the light curves are well-sampled both near and post peak, leading to better recovery of the input parameters. At higher redshifts, the majority of light curves are caught near peak and quickly drop below the detection limit, leading to typically poorer recovery. Additionally, due to the much deeper limits available in $gri$-bands, the light curves of higher redshift events are typically limited to these filters. Thus our ability to recover the input parameters significantly drops with redshift. At $z\lesssim 0.5$, our strict recovery rate is $\approx 60\%$ and our lenient recovery rate is $\approx 100\%$ for light curves with $>10$ data points.  By $z=2$, the strict recovery rate drops to zero, while the lenient recovery rate is $\approx 50$\%.  By $z=3$, the lenient recovery rate also drops to zero.

\begin{figure*}[t]
\includegraphics[width=0.95\textwidth]{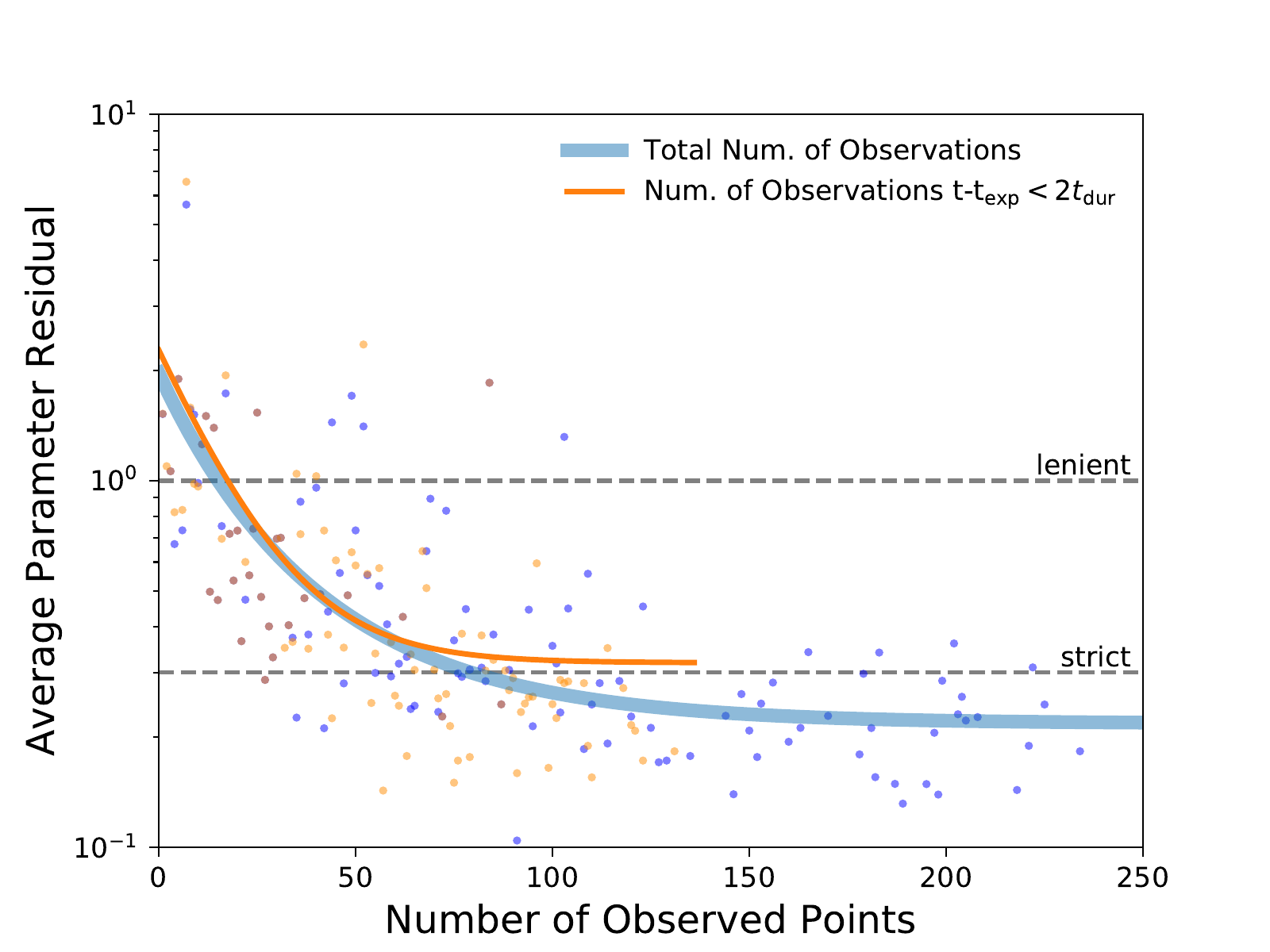}
\centering 
\caption{Average parameter residuals (for $M_\mathrm{ej},v_\mathrm{ej},B^*,P^*$) as a function of number of observations per light curve. The blue points show \textit{total} number of observations per light curve, while the oranges points show the number of observations within 100 days of explosion (in the event's rest frame). The solid lines are fits to exponential functions to guide the eye. Dotted lines show both the lenient and strict information criteria. There is little gain beyond $\approx 50$ observations per light curve, and almost no light curves have more than 50 observations within $2t_\mathrm{dur}$ days post-explosion.}
\label{fig:numptserr}
\end{figure*}

The parameter recovery rate is a function of both redshift and luminosity. The above calculations used a peak luminosity of $L_0\approx 3\times10^{44}$ erg s$^{-1}$. We now consider the full luminosity function of our simulated SLSNe (Figure~\ref{fig:lumfunc}) to capture the overall recovery rate. For any luminosity, the recovery curve as a function of redshift, $\epsilon_\mathrm{recov}(z)$, is set to 1 at $z=0$ and to zero at $z=z_\mathrm{lim}$, where $z=z_\mathrm{lim}$ is the limiting redshift for a given luminosity. For SLSNe with peak luminosity $L_0$, the limiting redshift is $z_\mathrm{lim,0}\approx 3.0$. We assume that for all other peak luminosities, the recovery rate can be described as $\epsilon_\mathrm{recov}=(z*z_\mathrm{lim,0}/z_\mathrm{lim})$. This allows brighter events to be captured at higher redshifts, and all events will be capped to their limiting redshifts. After reweighting the recovery rate through this process with our simulated luminosity function, we find that the overall efficiencies decline more rapidly with redshift. For example, the lenient recovery rate drops to $\approx 40$\% at $z=2$, rather than to $\approx 50$\% when we used just a single peak luminosity. 

Multiplying our corrected recovery rates by the overall discovery efficiency from \S\ref{sec:char}, we find that $\approx 18\%$ (about 1,700 out of 9,600 discovered annually) of SLSNe discovered in the WFD survey will have light curves that satisfy our strict criterion, and $\approx 50\%$ will satisfy the lenient criterion.  Even at the high redshift end ($z\approx 3$), $\approx 50$ SLSNe per year will satisfy the lenient criterion.  

For the DDFs, we find that the fraction of useful light curves is slightly smaller than that of the WFD survey. This is likely due to the fact that the overall efficiency reaches to higher redshifts, although fewer of the high-redshift light curves are useful due to their lower luminosities.

\subsection{Correlating SLSN Properties to Parameter Recovery}

Finally, we turn to the question of what properties of a SLSN light curve allow us to best recover key physical parameters. Unsurprisingly, the number of observations strongly correlates with our ability to recover parameters. This is demonstrated in Figure~\ref{fig:numptserr}, in which we show the average parameter residual for each of the four important physical parameters (e.g., $|M_\mathrm{ej,fit}-M_\mathrm{ej,true}|/M_\mathrm{ej,true}$) as a function of the number of observations for a sample of light curves spanning from $z=0.5$ to $z=2$. We consider both the total number of observations (taken at any point during the event) and observations taken within the first $2t_\mathrm{dur}$ days in the rest-frame (i.e., near-peak). In both cases, light curves with $\gtrsim 50$ points are significantly more likely to have recoverable physical parameters compared to the average light curve with $\approx 20$ observations ($\approx 65$\% compared to $\approx 50$\% recovery rate). Light curves with $\gtrsim 100$ observations are only somewhat more useful than those with $\gtrsim 50$ observations when using the lenient definition ($\approx70$\% vs $\approx65$\%); however, they are more useful when using the strict criterion, with $\approx40$\% compared to $\approx 30$\%. 

Additionally, having a measurable duration in any filter is a good indicator of an information-rich light curve, with $\approx65$\% compared to $\approx50$\% recovery rate for the typical light curve. This is likely due to the fact that the light curve peak and width greatly constrain the model parameter space.  For example, a bright and broad light curve cannot be produced by a small ejecta mass or weak magnetic field.  The most ``useful'' filter for measuring duration appears to be $u$-band, although this is likely due to the fact that light curves which are well-sampled in $u$-band tend to be at low redshifts (i.e., the limiting redshift for this metric is only $z=1.2$).  In contrast, light curves with a measurable $r$-band duration can occur at higher redshift ($z\approx 2$). Again, this suggests that well-sampled light curves near peak are more scientifically useful.

These findings indicate that a survey strategy which optimizes a higher cadence in the most sensitive bands, $gri$, will provide the greatest return on scientifically useful light curves even at high redshift. Given the average SLSN duration of $\approx 100$ days (Figure~\ref{fig:dlps}), a cadence of roughly two to four days (similar to the nominal cadence) in any filter would be sufficient to recover magnetar parameters directly from most SLSN light curves at $z<3.0$. Perhaps more importantly, the current WFD observing strategy has large seasonal gaps every $4-6$ months which interrupt many SLSN light curves. This is due to the fact that the WFD survey observes fields with airmass $\leq 1.4$ \citep{lsstsciencecollab}. Reducing these gaps with even a few observations at higher airmass would be beneficial to provide more comprehensive temporal coverage and greater opportunity to recover SLSN properties. In a similar vein, stacking late-time observations can significantly extend our light curve coverage accross seasons.

\section{Summary and Conclusions}
\label{sec:conclusions}

We presented detailed simulations of Type-I SLSNe in the upcoming LSST survey.  We constructed a realistic distribution of magnetar and explosion parameters from an existing sample of 58 SLSNe spanning $z=0.1-1.6$ and used this to simulate thousands of SLSNe at $z=0-6$ in the LSST Operations Simulator.

We define a number of measurable light curve metrics which we use to define a ``detection". For our loosest definition of a detection (observing $>10$ data points in all filters combined), we find that the detection efficiency of the WFD survey quickly declines from $\approx 50$\% at $z=1$ to $\approx 10$\% at $z=3$, while for the DDFs, the efficiency declines from $\approx 100$\% at $z=0.5$  to $\approx 50$\% at $z=3$ and $10\%$ at $z=5$. We combine this detection efficiency with an estimate for the cosmic SLSN rate to find that LSST will discover $\approx 10^4$ SLSNe per year within the WFD survey and $\approx 15$ per year in each DDF. Most ($90$\%) of the discovered SLSNe are found at $z\lesssim 3$, although $\approx 1$ SLSN per year should be discovered at $z\approx5$.

We refit the light curves of representative SLSNe injected into the LSST WFD survey and DDFs, and test how well we can recover four key physical parameters (initial magnetar spin period, magnetic field strength, ejecta velocity and ejecta mass). We find that we can successfully recover the four parameters in $\approx 18$\% of all SLSNe to within $30\%$ with error bars of $<50$\% of the parameter values. We can recover the parameters to within a factor of two for $\approx 50$\% of all SLSNe. The majority of SLSNe with recoverable parameters will be found at low redshift ($z\lesssim1.5$). Parameter recovery relies on having accurate redshifts; while LSST will provide photometric redshifts for many host galaxies this may become a challenge at the high redshift end.

We correlate our ability to recover physical parameters with the defined light curve metrics. In both the WFD survey and DDFs, light curves with $\gtrsim50$ observations, especially concentrated near-peak, are typically those with recoverable parameters. LSST survey strategies which maintain a rapid cadence ($\approx 2-4$ day) in the most sensitive $gri$ bands will provide the most scientifically useful SLSN light curves. Similarly, strategies which minimize seasonal gaps with some high airmass observations will increase our chance of covering the light curves peak and duration, and therefore provide more scientifically useful light curves. Finally, stacking observations at late times may allow us to probe more SLSNe across multiple seasons and better anchor our models. 

Compared to the WFD survey, we find that the DDFs (in their current form) will not provide higher quality SLSNe, or SLSNe at significantly higher redshifts in large quantities due to the small area covered by these fields. It is therefore imperative to maximize the scientific return from events in the WFD survey, rather than relying on a small number of events from the DDFs.

Overall, our simulations indicate that LSST will be a powerhouse for discovering SLSNe.  About 1,700 SLSNe per year will have sufficient photometry to extract key physical parameters directly from the light curves (given an accurate redshift estimate) to within 30\%,  significantly increasing our current sample by at least two orders of magnitude.

\acknowledgements
We thank J.~Guillochon and P.~Cowperthwaite for useful discussions. The computations presented in this work were performed on Harvard University’s Odyssey computer cluster, which is maintained by the Research Computing Group within the Faculty of Arts and Sciences. The Berger Time Domain group is supported in part by NSF grant AST-1714498 and NASA ADA grant NNX15AE50G.

\appendix
Using information provided by OpSim, we can calculate the signal-to-noise ratio ($SNR$) of our injected observations \footnote{See \url{https://smtn-002.lsst.io}}:
\begin{equation}
SNR = \frac{C}{\sqrt{C/g + (B/g + \sigma_\mathrm{instr}^2)*n_\mathrm{eff}}},
\end{equation}
where $C$ is the source counts in ADU, $B$ is the background count per pixel in ADU, $\sigma_\mathrm{instr}=12.7$ e$^-$ is the instrumental noise in ADU, $g=2.3$ e$^-$/ADU is the gain, and $n_\mathrm{eff}$ is the effective number of source pixels. Both $B$ and $\sigma_\mathrm{instr}$ are provided by {\tt OpSim}. The source counts are calculated using:

\begin{equation}
    C = \frac{A_\mathrm{eff}\Delta t}{g h}\int  F_\nu(\lambda)\frac{S(\lambda)}{ \lambda}d\lambda,
\end{equation}
where $F_\nu$ is source spectrum, and $S(\lambda)$ is the filter throughput, $\Delta t=30$s is the integration time, $A_\mathrm{eff}=3.24\times10^{10}$cm$^2$ is the effective collecting area and $h$ is Plank's constant. The effective number of pixels can be calculated as:

\begin{equation}
    n_\mathrm{eff} = 2.266(\text{FWHM}_\mathrm{eff}/\text{px})^2,
\end{equation}
where $\text{FWHM}_\mathrm{eff}$ is the effective full-width-at-half-max of the source PSF as reported by {\tt OpSim} and \text{px} $=0.2$''/pixel is the pixel scale.

For the DDFs, the exposure time is increased according to the number of exposures taken in a single night in each filter, allowing us to probe deeper limiting magnitudes.

\bigskip
\bibliography{mybib}{}

\end{document}